\documentclass[10pt,aps,twocolumn,secnumarabic,balancelastpage,amsmath,amssymb,nofootinbib,hyperref=pdftex,prx,citeautoscript,showpacs]{revtex4-1}

\usepackage{graphics}
\usepackage[pdftex]{graphicx}
\usepackage{tabularx}
\graphicspath{{./Figures/}}
\usepackage{longtable}
\usepackage{amsmath}
\usepackage{color,soul}

\makeatletter
\def\p@subsection{\thesection\,}
\makeatother

\newcommand{\bo}[1]{{\bf #1}}
\newcommand{\kk}{{\bf k}}

\newcommand{\x}{\bo{r}}

\usepackage{hyperref}
\hypersetup{colorlinks=true,linkcolor=blue,anchorcolor=blue,citecolor=blue,filecolor=blue,urlcolor=blue,bookmarksnumbered=true,pdfview=FitB}

\begin{document}

\title{Destructive interference of direct and crossed Andreev pairing in a system of two nanowires coupled via an $s$-wave superconductor}
\author{Christopher Reeg}
\author{Jelena Klinovaja}
\author{Daniel Loss}
\affiliation{Department of Physics, University of Basel, Klingelbergstrasse 82, CH-4056 Basel, Switzerland}
\date{\today}
\begin{abstract}
We consider a system of two one-dimensional nanowires coupled via an $s$-wave superconducting strip, a geometry that is capable of supporting Kramers pairs of Majorana fermions. By performing an exact analytical diagonalization of a tunneling Hamiltonian describing the proximity effect (via a Bogoliubov transformation), we show that the excitation gap of the system varies periodically on the scale of the Fermi wavelength in the limit where the interwire separation is shorter than the superconducting coherence length. Comparing with the excitation gaps in similar geometries containing only direct pairing, where one wire is decoupled from the superconductor, or only crossed Andreev pairing, where each nanowire is considered as a spin-polarized edge of a quantum Hall  state, we find that the gap is always reduced, by orders of magnitude in certain cases, when both types of pairing are present. Our analytical results are further supported by numerical calculations on a tight-binding lattice. Finally, we show that treating the proximity effect by integrating out the superconductor using the bulk Green's function does not reproduce the results of our exact diagonalization.
\end{abstract}

\maketitle

\paragraph*{Introduction.} Topological superconductivity has garnered a great deal of attention in recent years \cite{Alicea:2012,Leijnse:2012,Beenakker:2013} both theoretically  and experimentally because the localized excitations of such systems, known as Majorana fermions, obey non-Abelian statistics and can be utilized for applications in quantum computing \cite{Kitaev:2001,Nayak:2008}. The proposals which have received the most attention to date involve engineering Majorana states in nanowires with Rashba spin-orbit coupling in the presence of a Zeeman field \cite{Lutchyn:2010,Oreg:2010,Klinovaja:2012,Mourik:2012,Deng:2012,Das:2012,Rokhinson:2012,Churchill:2013,Finck:2013,Chang:2015,Albrecht:2016,Zhang:2016} or in ferromagnetic atomic chains \cite{NadjPerge:2014,Pawlak:2016,Klinovaja:2013,Vazifeh:2013,Braunecker:2013,NadjPerge:2013,Pientka:2013}. In the absence of any Zeeman splitting, it is possible to generate an even more exotic time-reversal invariant topological superconducting phase which supports Kramers pairs of Majorana fermions \cite{Wong:2012,Nakosai:2013,Zhang:2013,Keselman:2013,Schrade:2015,Haim:2014,Klinovaja:2014,Gaidamauskas:2014,Klinovaja:2014b,Danon:2015,Haim:2016}. One such proposal involves coupling two Rashba nanowires via an $s$-wave superconductor \cite{Klinovaja:2014,Gaidamauskas:2014}. In this system, superconductivity is induced in the nanowires via \emph{direct} Cooper pair tunneling, where both electrons of a Cooper pair tunnel into the same wire, and \emph{crossed Andreev} tunneling, where one electron from a Cooper pair tunnels into each wire \cite{Byers:1995,Choi:2000,Deutscher:2000,Lesovik:2001,Recher:2001,Yeyati:2007}. The topological phase can be realized if the strength of crossed Andreev pairing exceeds that of direct pairing. However, to date, the direct and crossed Andreev pairing strengths have been treated as theoretical parameters \cite{Klinovaja:2014,Gaidamauskas:2014,Klinovaja:2014b,Danon:2015} and no rigorous treatment of the proximity effect in this system has been carried out.

In this paper, we study the interplay between direct and crossed Andreev pairing in a noninteracting double-nanowire system by calculating the proximity-induced excitation gap as a function of the interwire separation ($d$). We show that the two pairing types always interfere destructively. When the tunneling strengths into each nanowire are equal, the excitation gap in the presence of both types of pairing is simply the difference between the gap in the presence of only direct pairing and the gap in the presence of only crossed Andreev pairing, with the direct gap always being larger than the crossed Andreev gap. When the interwire separation is shorter than the superconducting coherence length ($\xi_s$), this destructive interference can lead to an order of magnitude reduction in the size of the excitation gap of the system.

Our results are based on an exact analytical diagonalization of the tunneling Hamiltonian via a Bogoliubov transformation. We derive a set of effective Bogoliubov-de Gennes (BdG) equations that we then solve to determine the excitation gap as a function of $d$. Additionally, we show that integrating out the superconducting degrees of freedom using the bulk superconducting Green's function, a common method for treating the proximity effect in low-dimensional systems \cite{Alicea:2012,Sau:2010prox,Potter:2011,Stanescu:2011,Kopnin:2011,Zyuzin:2013,Takane:2014,vanHeck:2016,Reeg:2017}, yields incorrect and qualitatively different results in our finite geometry.

\begin{figure}[t!]
\includegraphics[width=.75\linewidth]{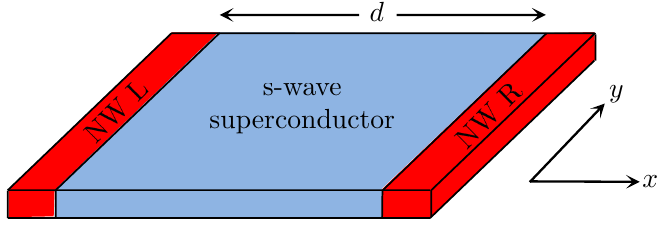}
\caption{\label{setup}Geometry of considered model. A 2D conventional $s$-wave superconductor of width $d$ ($0<x<d$) separates two 1D nanowires. The system is infinite in the $y$-direction.}
\end{figure}

\paragraph*{Model.} The system we consider is displayed in Fig.~\ref{setup}. Two one-dimensional nanowires are coupled to a superconducting strip of finite width $d$, taken to occupy $0<x<d$. The system is taken to be infinite in the $y$-direction, allowing us to define a conserved momentum $k_y$. We consider a Hamiltonian of the form
\begin{equation} \label{Hamiltonian}
H=H_{NW}^L+H_{NW}^R+H_{BCS}+H_t^L+H_t^R.
\end{equation}
The nanowire Hamiltonian can be expressed generally as
\begin{equation} \label{Hnw}
H_{NW}^i=\sum_{\sigma,\sigma'}\int\frac{dk_y}{2\pi}\,\psi^\dagger_{i\sigma}(k_y)\mathcal{H}_{\sigma\sigma'}^i(k_y)\psi_{i\sigma'}(k_y),
\end{equation}
where $\psi_{i\sigma}^\dagger(k_y)$ [$\psi_{i\sigma}(k_y)$] creates (annihilates) an electron of spin $\sigma$ and momentum $k_y$ in nanowire $i$ and the Hamiltonian density ${\mathcal{H}}_i(k_y)$ of each wire is left unspecified. We describe the superconductor by a BCS Hamiltonian,
\begin{equation} \label{Hs}
\begin{aligned}
H_{BCS}=\int\frac{dk_y}{2\pi}\int dx\,\Psi^\dagger_s(H_0\tau_z+\Delta\tau_x)\Psi_s,
\end{aligned}
\end{equation}
where $\Psi_s=[\psi_{s\uparrow}(-k_y,x),\psi_{s\downarrow}^\dagger(k_y,x)]^T$, $\psi_{s\sigma}^\dagger(k_y,x)$ [$\psi_{s\sigma}(k_y,x)$] creates (annihilates) an electron of spin $\sigma$ and momentum $k_y$ at position $x$ inside the superconductor, $H_0=-\partial_x^2/2m_s+k_y^2/2m_s-\mu_s$ ($m_s$ is the effective mass and $\mu_s$ is the Fermi energy of the superconductor), $\Delta$ is the superconducting pairing potential, and $\tau_{x,y,z}$ are Pauli matrices acting in Nambu space. We also allow for electrons to tunnel between superconductor and wire, assuming that this process preserves both spin and momentum. Tunneling is described by
\begin{equation} \label{Ht}
H_t^i=-t_i\sum_\sigma \int\frac{dk_y}{2\pi}\left[\psi_{i\sigma}^\dagger(k_y)\psi_{s\sigma}(k_y,x_i)+H.c.\right],
\end{equation}
where $t_i$ is a wire-dependent tunneling amplitude and $x_i$ denotes the position of wire $i$.

\paragraph*{Bogoliubov-de Gennes Equations.} To solve the model under consideration, we perform an exact diagonalization of Hamiltonian (\ref{Hamiltonian}) by introducing a transformation of the form
\begin{subequations} \label{Bogoliubov}
\begin{gather}
\psi_{i\sigma}^\dagger(k_y)=\sum_n\left[\gamma_n^\dagger u_{in\sigma}^*(k_y)+\gamma_nv_{in\sigma}(k_y)\right], \\
\psi_{s\sigma}^\dagger(k_y,x)=\sum_n\left[\gamma_n^\dagger u_{sn\sigma}^*(k_y,x)+\gamma_nv_{sn\sigma}(k_y,x)\right],
\end{gather}
\end{subequations}
where $\gamma_n$ describes the new quasiparticles of the system and $u(v)$ are electron (hole) wave functions. It is straightforward to show \cite{supp} that transformation (\ref{Bogoliubov}) diagonalizes Hamiltonian (\ref{Hamiltonian}) provided that the wave functions obey a set of BdG equations given by
\begin{subequations} \label{BdG}
\begin{align}
Eu_{i}(k_y)&={\mathcal{H}}_i(k_y)u_{i}(k_y)-t_iu_{s}(k_y,x_i), \label{BdG1} \\
-Ev_{i}(k_y)&={\mathcal{H}}^T_i(k_y)v_{i}(k_y)-t_iv_{s}(k_y,x_i), \label{BdG2} \\
Eu_{s}(k_y,x)&=H_0u_{s}(k_y,x)+\Delta(i\sigma_y)v_{s}(-k_y,x) \label{BdG3} \\
	&\hspace*{0.3in}-\sum_i t_i\delta(x-x_i)u_{i}(k_y), \nonumber \\
-Ev_{s}(k_y,x)&=H_0v_{s}(k_y,x)+\Delta(i\sigma_y) u_{s}(-k_y,x) \label{BdG4} \\
	&\hspace*{0.3in}-\sum_i t_i\delta(x-x_i)v_{i}(k_y). \nonumber
\end{align}
\end{subequations}
In Eqs.~(\ref{BdG}), we introduce the spinor electron (hole) wave function $u(v)_{j}=[u(v)_{j\uparrow},u(v)_{j\downarrow}]^T$ for $j=i,s$ and denote the Pauli matrix acting in spin space by $\sigma_{x,y,z}$. 

While Eqs.~(\ref{BdG}) were derived without making any assumptions about the nanowire Hamiltonian, for the remainder of the paper we focus on the simple case where each nanowire is a normal conductor that can be described by ${\mathcal{H}}_i(k_y)=\xi_k$, with $\xi_k=k_y^2/2m_n-\mu_n$ ($m_n$ and $\mu_n$ are the effective mass and Fermi energy of the nanowires). With this simple choice for the nanowire Hamiltonian, we are able to eliminate the trivial spin sector from the BdG equations; essentially, we can reduce Eqs.~(\ref{BdG}) from matrix equations to scalar equations. Equations~(\ref{BdG1}) and (\ref{BdG2}) form an independent algebraic system that yields the solutions
\begin{equation} \label{BdGsol1}
u_{i\uparrow}[v_{i\downarrow}](k_y)=\frac{t_i}{\xi_k\mp E}u_{s\uparrow}[v_{s\downarrow}](k_y,x_i).
\end{equation}
Substituting Eq.~(\ref{BdGsol1}), we can decouple Eqs.~(\ref{BdG3}) and (\ref{BdG4}) to obtain a system of differential equations describing the wave functions in the superconductor,
\begin{equation} \label{eqs}
\begin{aligned}
&\biggl(\pm H_0+\frac{t_L^2\delta(x-x_L)}{E\mp\xi_k}+\frac{t_R^2\delta(x-x_R)}{E\mp\xi_k}\biggr)u_{s\uparrow}[v_{s\downarrow}](k_y,x) \\
	&\hspace*{0.3in}+\Delta v_{s\downarrow}[u_{s\uparrow}](-k_y,x)=Eu_{s\uparrow}[v_{s\downarrow}](k_y,x),
\end{aligned}
\end{equation}
The solution to Eq.~{\eqref{eqs}} within the left $l$ ($0<x<x_L$), middle $m$ ($x_L<x<x_R$) and right $r$ ($x_R<x<d$) regions of the superconductor is
\begin{equation}
\begin{aligned}
\psi_l(k_y,x)&=c_1\begin{pmatrix} u_0 \\ v_0 \end{pmatrix}\sin(p_+x)+c_2\begin{pmatrix} v_0 \\ u_0 \end{pmatrix}\sin(p_-x), \\
\psi_m(k_y,x)&= c_3\begin{pmatrix} u_0 \\ v_0 \end{pmatrix}e^{ip_+x}+c_4\begin{pmatrix} u_0 \\ v_0 \end{pmatrix}e^{-ip_+x} \\
	&+c_5\begin{pmatrix} v_0 \\ u_0 \end{pmatrix}e^{ip_-x}+c_6\begin{pmatrix} v_0 \\ u_0 \end{pmatrix}e^{-ip_-x}, \\
\psi_r(k_y,x)&=c_7\begin{pmatrix} u_0 \\ v_0 \end{pmatrix}\sin[p_+(d-x)]+c_8\begin{pmatrix} v_0 \\ u_0 \end{pmatrix}\sin[p_-(d-x)],
\end{aligned}
\end{equation}
where $\psi(k_y,x)=[u_{s\uparrow}(k_y,x),v_{s\downarrow}(-k_y,x)]^T$ is a spinor wave function, $p_\pm^2=2m_s(\mu_s\pm i\Omega)-k_y^2$, $\Omega^2=\Delta^2-E^2$, and $u_0^2(v_0^2)=(1\pm i\Omega/E)/2$. To determine the eight unknown coefficients $c_{1-8}$, we must impose appropriate boundary conditions at $x=x_L$ and $x=x_R$ (note that a vanishing boundary condition at $x=0$ and $x=d$ has already been imposed). In addition to continuity of the wave function, the boundary conditions on the derivatives of the wave functions are determined by the delta-function terms of Eqs.~(\ref{eqs}) and are obtained by direct integration:
\begin{subequations} \label{BCs}
\begin{align}
\partial_xu_{s\uparrow}[v_{s\downarrow}](k_y,x_L)&=\pm\frac{2k_{F}\gamma_L}{E\mp\xi_k}u_{s\uparrow}[v_{s\downarrow}](k_y,x_L), \\
\partial_xu_{s\uparrow}[v_{s\downarrow}](k_y,x_R)&=\mp\frac{2k_{F}\gamma_R}{E\mp\xi_k}u_{s\uparrow}[v_{s\downarrow}](k_y,x_R).
\end{align}
\end{subequations}
In Eqs.~(\ref{BCs}) we introduce an energy scale associated with tunneling which is proportional to the square of the tunneling amplitude, $\gamma_i=t_i^2/v_F$, where $v_{F}=k_F/m_s$ is the Fermi velocity of the superconductor. Assuming that the Fermi momentum of the superconductor greatly exceeds that of the nanowires ($k_{F}\gg k_{Fn}$) allows us to approximate $p_\pm=k_{F}\pm i\Omega/v_{F}$ (because $k_y$ is conserved, typical values take $k_y\lesssim k_{Fn}\ll k_{F}$; we also expand in the limit $\mu_s\gg\Delta$). However, even by making these simplifications the solvability condition of Eqs.~(\ref{BCs}) cannot be solved besides numerically for an arbitrary parameter set \cite{supp}.

\begin{figure}[t!]
\includegraphics[width=\linewidth]{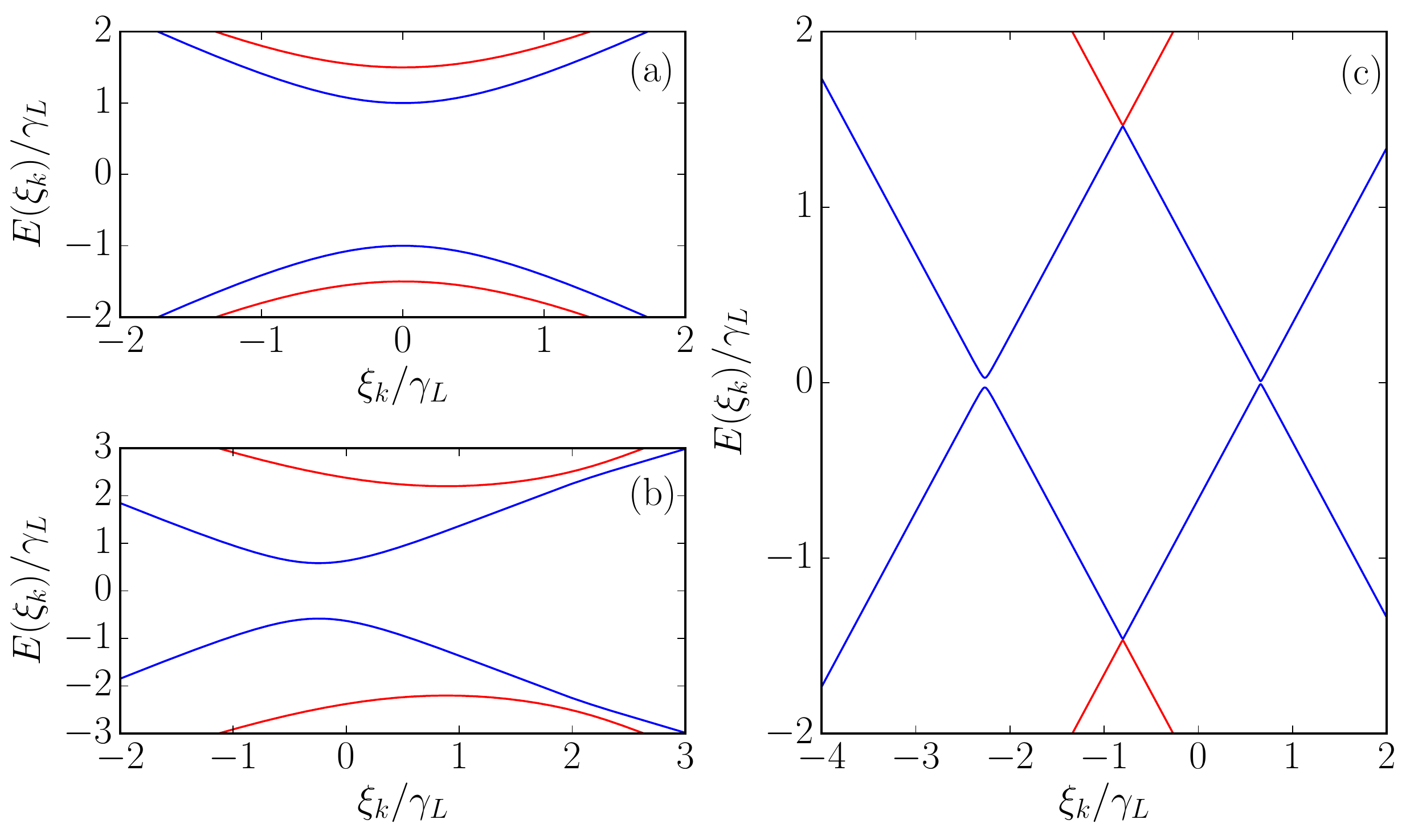}
\caption{\label{spectra}Excitation spectra for (a) $d=100\xi_s$, (b) $d=\xi_s$, and (c) $d=0.01\xi_s$. For all plots, we choose $\gamma_R/\gamma_L=1.5$ and $k_F\xi_s=100$.}
\end{figure}

In order to proceed analytically, we assume that the superconductor is only weakly coupled to each nanowire, so that $\gamma_i\ll\Delta$. In this limit, the relevant pairing energies in the nanowires are small and we can focus our attention on energies $E\ll\Delta$.  We also assume that the nanowires are (symmetrically) located near the ends of the superconductor, such that $x_L=x_w$ and $x_R=d-x_w$ with $x_w\ll d$. The solvability condition in this limit can be expressed as $a(\xi_k,d)E^4-b(\xi_k,d)E^2+c(\xi_k,d)=0$,
with the complicated expressions for the coefficients given in the Supplementary Material \cite{supp}. This equation can be solved exactly for the energy spectrum $E(\xi_k)$, which is plotted for several values of $d$ in Fig.~\ref{spectra}. When $d\gg\xi_s$ [Fig.~\ref{spectra}(a)], the spectrum consists of two parabolic bands and has a gap given by $\min\{\gamma_L,\gamma_R\}$; this represents the decoupling of two nanowires with a large spatial separation. When the wires are brought closer together [Fig.~\ref{spectra}(b-c)], crossed Andreev pairing reduces the size of the gap while single-particle couplings induced by tunneling effectively shift the chemical potentials of each band \cite{Klinovaja:2014b,supp}.

\paragraph*{Excitation gap in the weak-coupling limit.} 
Our goal is to calculate the excitation gap (the global minimum of the spectrum) as a function of $d$. Although we are able to solve for the spectrum exactly, it is still quite cumbersome to determine the excitation gap for all $d$ when tunneling is asymmetric ($\gamma_L\neq\gamma_R$).

Let us consider the symmetric-tunneling case $\gamma_L=\gamma_R$, leaving the asymmetric case for the supplemental material \cite{supp}. Under the assumption of symmetric tunneling, it is quite straightforward to solve for the gap for any value of $d$ \cite{supp}. Assuming that $d/\xi_s\gg\gamma/\Delta$, the gap is
\begin{equation} \label{gapsymm}
E_g(d)=\frac{\gamma\sinh(d/\xi_s)}{\cosh(d/\xi_s)+|\cos (k_Fd)|}.
\end{equation}
[The gap in principle also depends on the wire position $x_w$ through an additional factor $\sin^2(k_Fx_w)$; because this is a rather arbitrary quantity, we simply replace it by its mean value $\langle\sin^2(k_Fx_w)\rangle=1/2$ throughout.] When the superconductor is very wide ($d\gg\xi_s$), the gap approaches $E_g=\gamma$. When the superconductor is very narrow ($d\ll\xi_s$), the gap oscillates on the scale of $1/k_F$ between its maximum value $E_g^\text{max}=\gamma d/\xi_s$, attained for $k_Fd=\pi(n+1/2)$ ($n\in\mathbb{Z}$), and its minimum value $E_g^\text{min}=\gamma d/2\xi_s$, attained for $k_Fd=n\pi$.

Note that because we chose $\gamma_L=\gamma_R$, we are unable to distinguish between direct and crossed Andreev pairing in our result for the gap [Eq.~(\ref{gapsymm})]. We again must stress that we are not solving an effective model, so the direct and crossed Andreev pairing functions are not parameters of our theory as in Refs.~\cite{Gaidamauskas:2014,Klinovaja:2014,Klinovaja:2014b,Danon:2015}. Instead, we identify direct terms as being proportional to $t_i^2$ ($\gamma_i$) and crossed Andreev terms as being proportional or $t_Lt_R$ ($\sqrt{\gamma_L\gamma_R}$). In an attempt to differentiate between the two contributions, we compare the gap in the presence of both pairing types to that of similar systems containing only one type of pairing.

First, we isolate direct pairing in our system by decoupling one of the wires from the superconductor. Setting $t_L=0$ in Eq.~(\ref{eqs}), we find a direct gap of the form \cite{supp}
\begin{equation} \label{directgap}
E_g^D(d)=\frac{\gamma\sinh(2d/\xi_s)}{\cosh(2d/\xi_s)-\cos (2k_Fd)}.
\end{equation}
If the superconductor is very wide, the gap approaches $E_g=\gamma$ as before. If the superconductor is very narrow, the gap is $E_g=(2\gamma d/\xi_s)/[1-\cos (2k_Fd)+2d^2/\xi_s^2]$. The gap is sharply peaked near $k_Fd=n\pi$ and has a maximum value $E_g^{D,\text{max}}=\gamma\xi_s/d$. The gap is minimized near $k_Fd=\pi(n+1/2)$ and takes the value $E_g^{D,\text{min}}=\gamma d/\xi_s$.

\begin{figure*}[t!]
\includegraphics[width=\linewidth]{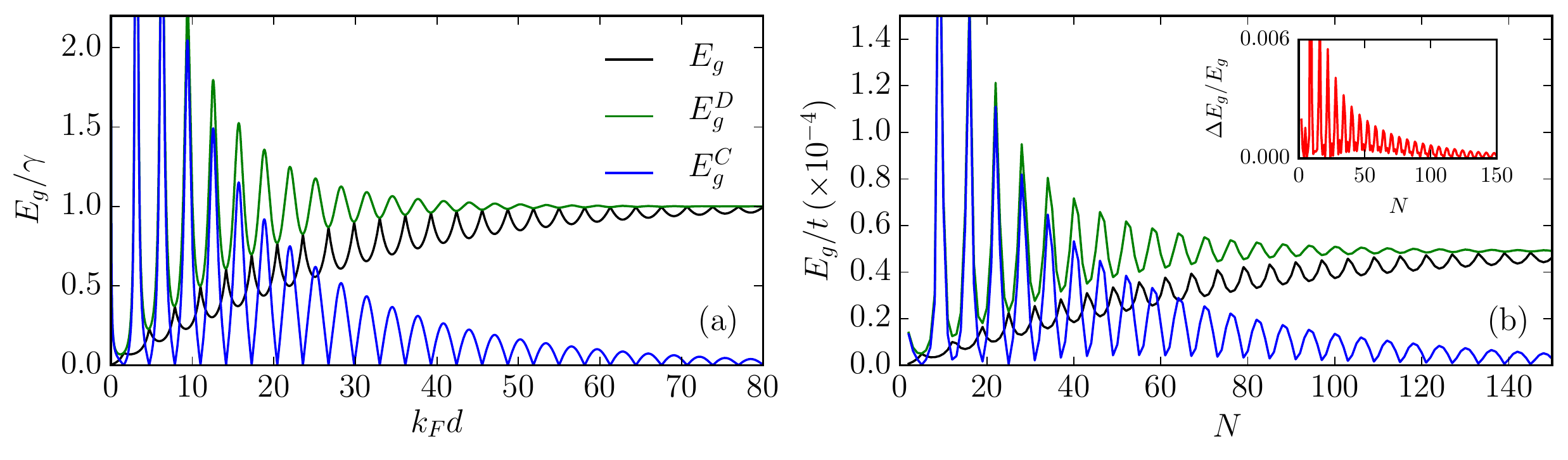}
\caption{\label{gapvswidth} Proximity-induced gaps plotted as a function of superconductor width. (a) Analytical results for $k_F\xi_s=20$. Black curve corresponds to symmetric tunneling ($\gamma_L=\gamma_R$) [Eq.~(\ref{gapsymm})], green curve corresponds to case of a single wire ($\gamma_L=0$) [Eq.~(\ref{directgap})], and blue curve corresponds to quantum Hall regime ($\gamma_{L\uparrow}=\gamma_{R\downarrow}\neq0$ and $\gamma_{L\downarrow}=\gamma_{R\uparrow}=0$) [Eq.~(\ref{crossedgap})] (b) Numerical results for $\Delta=0.02t$, $\mu_s=0.3t$, $\mu_n=0.03t$, and $t_L=t_R=0.01t$, where $t$ is the hopping parameter in the superconductor. Inset: plot of $\Delta E_g\equiv |E_g-(E_g^D-E_g^C)|$, showing very good quantitative agreement with Eq.~(\ref{relationship}).}
\end{figure*}

To isolate crossed Andreev pairing in our system, we consider a situation where both nanowires are spin-polarized and have opposite spin; i.e., they are edge states of two quantum Hall systems with opposite chirality. In this case, we introduce a spin dependence to the tunneling amplitudes, $t_i\to t_{i\sigma}$. Assuming for example that $t_{L\uparrow}=t_{R\downarrow}\neq0$ while $t_{L\downarrow}=t_{R\uparrow}=0$, we set $\gamma_R=0$ in the equation for the electron wave function (which has spin-up) and $\gamma_L=0$ in the equation for the hole wave function (which has spin-down) to find a crossed Andreev gap given by \cite{supp}
\begin{equation} \label{crossedgap}
E_g^C(d)=\frac{2\gamma\sinh(d/\xi_s)}{\cosh(2d/\xi_s)-\cos (2k_Fd)}|\cos (k_Fd)|.
\end{equation}
If the superconductor is very wide, the gap oscillates on the scale $1/k_F$ and decays on the scale $\xi_s$, $E_g=2\gamma|\cos (k_Fd)|e^{-d/\xi_s}$. If the superconductor is very narrow, we expand to find $E_g=(2\gamma d/\xi_s)|\cos (k_Fd)|/(1-\cos (2k_Fd)+2d^2/\xi_s^2)$. Similarly to the direct pairing case, the gap is sharply peaked near $k_Fd=n\pi$, having a maximum value $E_g^{C,\text{max}}=\gamma\xi_s/d$. The crossed Andreev gap is minimized near $k_Fd=\pi(n+1/2)$, where it vanishes. The vanishing of the gap indicates a change in sign of the crossed Andreev pairing function (see also \cite{supp}). Therefore, it should be possible to form a $\pi$-junction by coupling two systems of different $d$. Such $\pi$ phase shifts are crucial for engineering Majorana fermions in similar setups \cite{Keselman:2013,Schrade:2015}.

The three gaps that we have calculated are plotted in Fig.~\ref{gapvswidth}(a). The gaps are related through the remarkably simple expression
\begin{equation} \label{relationship}
E_g(d)=E_g^D(d)-E_g^C(d),
\end{equation}
indicating that direct and crossed Andreev pairing interfere with one another destructively. This effect is maximized when the superconductor is very narrow, as crossed Andreev reflection is not significantly suppressed by the interwire separation. Quite interestingly, because the direct and crossed Andreev gaps attain their maxima at the same thickness ($k_Fd=n\pi$), the gap $E_g$ is minimized when pairing is maximized. Furthermore, destructive interference between the two pairing processes at these points leads to an order of magnitude reduction of the gap [specifically, a reduction by a factor of order $\mathcal{O}(\xi_s^2/d^2)$].

We also support our analytical results with a standard tight-binding calculation in the geometry of Fig.~\ref{setup} \cite{supp}. Results of the tight-binding calculation are plotted in Fig.~\ref{gapvswidth}(b), showing very good qualitative agreement with Fig.~\ref{gapvswidth}(a). We also plot the quantity $\Delta E_g\equiv|E_g-(E_g^D-E_g^C)|$ in the inset of Fig.~\ref{gapvswidth}(b), showing very good quantitative agreement with Eq.~(\ref{relationship}).

\paragraph*{Integrating out superconductor.} Finally, we show that integrating out the superconducting degrees of freedom from Eq.~(\ref{Hamiltonian}) using the bulk superconducting Green's function does not reproduce the results of our exact diagonalization in a finite geometry. Assuming that tunneling is weak and symmetric ($\gamma_L=\gamma_R\ll\Delta$) and that the superconductor is very narrow ($d\ll\xi_s$), we integrate out the superconductor to yield an effective Hamiltonian describing superconductivity induced in the nanowires \cite{supp}. This effective Hamiltonian yields a low-energy spectrum $E_\pm^2(k)=\delta^2\gamma_c^2+(\beta\gamma_d+\xi_k\pm\eta\gamma_c)^2$, where $\gamma_{d(c)}$ differentiate one-wire (two-wire) tunneling processes, $\beta=\cot (k_Fd/2)$, $\delta=-\cos (k_Fd/2)$, and $\eta=\cos (k_Fd/2)\cot (k_Fd/2)$. In obtaining the low-energy spectrum, we expanded the effective Hamiltonian to order $\mathcal{O}[(d/\xi_s)^0]$. The minimum excitation gap of the spectrum is $E_{g}=\gamma_c|\cos (k_Fd/2)|$. Therefore, if the superconductor is integrated out using the bulk Green's function, one would incorrectly find that crossed Andreev pairing always dominates over direct pairing in the limit $d\ll\xi_s$ [note that direct pairing shows up in the effective Hamiltonian only at order $\mathcal{O}(d/\xi_s)$]. 

Physically, this procedure gives a false result because it fails to properly account for the boundary conditions that must be imposed when evaluating the Gaussian path integral. When the width of the superconductor is small compared to the coherence length, these boundary effects cannot be neglected. We find that integrating out the superconductor using the bulk Green's function reproduces the correct spectrum only in the limit $d\gg\xi_s$, when the boundary effects can be safely neglected \cite{supp}.We also discuss in the supplemental material how one can properly account for the boundary effects when integrating out \cite{supp}.

\paragraph*{Conclusions.} We have shown that direct and crossed Andreev pairing interfere destructively in a system of two nanowires coupled via an $s$-wave superconducting strip. When the interwire separation $d$ is shorter than the coherence length $\xi_s$, this destructive interference can lead to an order of magnitude reduction in the size of the excitation gap when compared to similar systems containing only a single type of pairing. Our analytical solution is based on an exact treatment of the proximity effect through the diagonalization of the tunneling Hamiltonian (via a Bogoliubov transformation) and is supported by numerical tight-binding calculations. Additionally, we have explicitly shown that integrating out the superconductor using the bulk Green's function does not reproduce the results of our exact diagonalization.

The interference effects discussed in this paper, which are manifested through oscillations of the excitation gap on the scale of the Fermi wavelength $1/k_F$, can most easily be observed when the interwire separation is smaller than the coherence length $\xi_s$. If the superconductor is metallic, observing these oscillations is not feasible. However, proximity-inducing superconductivity in a low-density semiconducting two-dimensional electron gas (2DEG) such as InGaAs/InAs (as in \cite{Kjaergaard:2016}), has several advantages. Inducing superconductivity by the proximity effect will make both $\xi_s$ and $1/k_F$ larger and will allow the density of the 2DEG to be tuned with a gate voltage (so that $k_Fd$ can be varied using a single sample). Due to our assumption of translational invariance along the $y$-direction, the interface between superconductor and nanowire must be made smooth (on the scale of $\xi_s$) and $d$ must be made uniform.

Finally, we note that crossed Andreev pairing is always weaker than direct pairing in the absence of interactions. Therefore, intrawire repulsive electron-electron interactions are needed to stabilize the time-reversal-invariant topological phase in the double-nanowire system, as they can significantly reduce direct pairing while leaving crossed Andreev pairing unaffected \cite{Recher:2002,Bena:2002}. In this case, the nanowires support Kramers pairs of Majorana fermions and parafermions \cite{Klinovaja:2014}. However, the destructive interference between direct and crossed Andreev pairing in the double-wire setup allows for a conventional topological superconducting phase to form at significantly reduced magnetic field strengths compared to the case of a single wire with only direct pairing \cite{Schrade:2017}.

\paragraph*{Acknowledgments.}
We thank C. Schrade and Y. Volpez for useful discussions. This work was supported by the Swiss National Science Foundation and the NCCR QSIT.

\bibliography{bibCrossedAndreev}

\newpage

\begin{widetext}

\setcounter{equation}{0}
\setcounter{figure}{0}

\section*{Supplemental Material}

\section{Bogoliubov Transformation to Diagonalize Tunneling Hamiltonian} \label{Sec1supp}
We consider a tunneling Hamiltonian which couples an $s$-wave superconductor to two nanowires. Such a Hamiltonian can be written generally as
\begin{equation} \label{Hamiltoniansupp}
H=H_{NW}^L+H_{NW}^R+H_{BCS}+H_t^L+H_t^R.
\end{equation}
For now, we leave the nanowire Hamiltonian unspecified,
\begin{equation} \label{HNWsupp}
H_{NW}^i=\frac{1}{2}\sum_{\sigma,\sigma'}\int dy_i\left[\psi_{i\sigma}^\dagger(y_i)\mathcal{H}_{\sigma\sigma'}^i(y_i)\psi_{i\sigma'}(y_i)-\psi_{i\sigma}(y_i)\mathcal{H}_{\sigma\sigma'}^{iT}(y_i)\psi^\dagger_{i\sigma'}(y_i)\right],
\end{equation}
where $y_i$ denotes the coordinate in wire $i$ and $\mathcal{H}^i_{\sigma\sigma}(y_i)$ is the Hamiltonian of wire $i$. We describe the superconductor using a conventional BCS Hamiltonian given by
\begin{equation} \label{HBCSsupp}
\begin{aligned}
H_{BCS}&=\frac{1}{2}\sum_{\sigma,\sigma'}\int d\x\biggl\{\psi_{s\sigma}^\dagger(\x)H_0(\x)\psi_{s\sigma}(\x)-\psi_{s\sigma}(\x)H_0(\x)\psi^\dagger_{s\sigma}(\x)+\psi_{s\sigma}^\dagger(\x)\Delta_{\sigma\sigma'}\psi_{s\sigma'}^\dagger(\x)+\psi_{s\sigma}(\x)\Delta_{\sigma\sigma'}^\dagger\psi_{s\sigma'}(\x)\biggr\},
\end{aligned}
\end{equation}
where $\x=(x,y)$ denotes the coordinate in the superconductor. We define the (constant in space) pairing potential matrix by $\Delta_{\sigma\sigma'}=\Delta (i\sigma_y)_{\sigma\sigma'}$ and $H_0(\x)=-\nabla^2/2m_s-\mu_s$ ($m_s$ and $\mu_s$ are the effective mass and Fermi energy of the superconductor, respectively). Finally, we have allowed for tunneling between the superconductor and each wire; such a tunneling Hamiltonian can be expressed generally as
\begin{equation} \label{Htsupp}
\begin{aligned}
H_t^{i}&=\frac{1}{2}\sum_{\sigma,\sigma'}\int dy_i\int d\x\biggl\{\psi_{s\sigma}^\dagger(\x)T_{\sigma\sigma'}^i(\x,y_i)\psi_{i\sigma}(y_i)+\psi^\dagger_{i\sigma}(y_i)T^{i\dagger}_{\sigma\sigma'}(y_i,\x)\psi_{s\sigma}(\x) \\
	&-\psi_{s\sigma}(\x)T_{\sigma\sigma'}^{i*}(\x,y_i)\psi_{i\sigma}^\dagger(y_i)-\psi_{i\sigma}(y_i)T^{iT}_{\sigma\sigma'}(y_i,\x)\psi_{s\sigma}^\dagger(\x)\biggr\},
\end{aligned}
\end{equation}
where $T_{\sigma\sigma'}^i(\x,y_i)$ is a tunneling matrix element coupling a state in the superconductor of spin $\sigma$ at position $\x$ to a state in wire $i$ of spin $\sigma'$ at position $y_i$. The matrix element also satisfies the identity $T_{\sigma'\sigma}^{i*}(\x,y_i)=T_{\sigma\sigma'}^{i\dagger}(y_i,\x)$. To diagonalize Hamiltonian (\ref{Hamiltoniansupp}), we introduce a Bogoliubov transformation of the form
\begin{equation} \label{Bogoliubovsupp}
\begin{gathered}
\psi_{i\sigma}(y_i)=\sum_n\left[\gamma_n u_{in\sigma}(y_i)+\gamma_n^\dagger v_{in\sigma}^*(y_i)\right], \\
\psi_{s\sigma}(\x)=\sum_n\left[\gamma_n u_{sn\sigma}(\x)+\gamma_n^\dagger v_{sn\sigma}^*(\x)\right].
\end{gathered}
\end{equation}
To ensure appropriate fermionic commutation relations for the quasiparticle operators $\gamma_n$, the wave functions in the nanowires must obey the following relations:
\begin{equation} \label{relationsupp1}
\begin{gathered}
\sum_{n}\left[u_{in\sigma}^*(y_i)u_{jn\sigma'}(y_i')+v_{in\sigma}(y_i)v^*_{jn\sigma'}(y_i')\right]=\delta_{ij}\delta_{\sigma\sigma'}\delta(y_i-y_i'), \\
\sum_{n}\left[u_{in\sigma}^*(y_i)v_{jn\sigma'}(y_i')+v_{in\sigma}(y_i)u^*_{jn\sigma'}(y_i')\right]=0.
\end{gathered}
\end{equation}
Similarly, the wave functions in the superconductor must obey
\begin{equation} \label{relationsupp2}
\begin{gathered}
\sum_{n}\left[u_{sn\sigma}^*(\x)u_{sn\sigma'}(\x')+v_{sn\sigma}(\x)v^*_{sn\sigma'}(\x')\right]=\delta_{\sigma\sigma'}\delta(\x-\x'), \\
\sum_{n}\left[u_{sn\sigma}^*(\x)v_{sn\sigma'}(\x')+v_{sn\sigma}(\x)u^*_{sn\sigma'}(\x')\right]=0.
\end{gathered}
\end{equation}
Additionally, the product of nanowire and superconducting wave functions must obey
\begin{equation} \label{relationsupp3}
\begin{gathered}
\sum_{n}\left[u^*_{in\sigma}(y_i)u_{sn\sigma'}(\x')+v_{in\sigma}(y_i)v^*_{sn\sigma'}(\x')\right]=0, \\
\sum_{n}\left[u^*_{in\sigma}(y_i)v_{sn\sigma'}(\x')+v_{in\sigma}(y_i)u^*_{sn\sigma'}(\x')\right]=0.
\end{gathered}
\end{equation}
Inverting Eqs.~(\ref{Bogoliubovsupp}), the quasiparticle operators $\gamma_n$ can be expressed in terms of the nanowire and superconductor fermion operators as
\begin{equation}
\gamma_n=\sum_{\sigma}\biggl\{\sum_i\int dy_i\left[u^*_{in\sigma}(y_i)\psi_{i\sigma}(y_i)+v^*_{in\sigma}(y_i)\psi^\dagger_{i\sigma}(y_i)\right]+\int d\x\left[u^*_{sn\sigma}(\x)\psi_{s\sigma}(\x)+v^*_{sn\sigma}(\x)\psi^\dagger_{s\sigma}(\x)\right]\biggr\}.
\end{equation} 
Inverting also the relations in Eqs.~\eqref{relationsupp1}-\eqref{relationsupp3}, we obtain two additional constraints on the wave functions:
\begin{equation} \label{relationssupp}
\begin{gathered}
\sum_{\sigma}\biggl\{\sum_i\int dy_i\left[u_{im\sigma}(y_i)u^*_{in\sigma}(y_i)+v_{im\sigma}(y_i)v^*_{in\sigma}(y_i)\right]+\int d\x\left[u_{sm\sigma}(\x)u^*_{sn\sigma}(\x)+v_{sm\sigma}(\x)v^*_{sn\sigma}(\x)\right]\biggr\}=\delta_{mn}, \\
\sum_{\sigma}\biggl\{\sum_i\int dy_i\left[u_{im\sigma}(y_i)v_{in\sigma}(y_i)+v_{im\sigma}(y_i)u_{in\sigma}(y_i)\right] +\int d\x\left[u_{sm\sigma}(\x)v_{sn\sigma}(\x)+v_{sm\sigma}(\x)u_{sn\sigma}(\x)\right]\biggr\}=0.
\end{gathered}
\end{equation}
We now substitute transformation (\ref{Bogoliubovsupp}) into Eq.~(\ref{Hamiltoniansupp}) and define the Bogoliubov-de Gennes (BdG) equations by
\begin{equation}
\begin{gathered}
\sum_{\sigma'}\left[\mathcal{H}^i_{\sigma\sigma'}(y_i)u_{in\sigma'}(y_i)+\int d\x\,T^{i\dagger}_{\sigma\sigma'}(y_i,\x)u_{sn\sigma}(\x)\right]=E_nu_{in\sigma}(y_i), \\
\sum_{\sigma'}\left[-\mathcal{H}^{iT}_{\sigma\sigma'}(y_i)v_{in\sigma'}(y_i)-\int d\x\,T^{iT}_{\sigma\sigma'}(y_i,\x)v_{sn\sigma}(\x)\right]=E_nv_{in\sigma}(y_i), \\
\sum_{\sigma'}\left[H_0(\x)u_{sn\sigma}(\x)+\Delta_{\sigma\sigma'}v_{sn\sigma'}(\x)+\sum_i\int dy_i\,T^i_{\sigma\sigma'}(\x,y_i)u_{in\sigma}(y_i)\right]=E_nu_{sn\sigma}(\x), \\
\sum_{\sigma'}\left[-H_0(\x)v_{sn\sigma}(\x)+\Delta^\dagger_{\sigma\sigma'}u_{sn\sigma'}(\x)-\sum_i\int dy_i\,T^{i*}_{\sigma\sigma'}(\x,y_i)v_{in\sigma}(y_i)\right]=E_nv_{sn\sigma}(\x). 
\end{gathered}
\end{equation}
After defining the BdG equations, the Hamiltonian can be expressed as
\begin{equation}
\begin{aligned}
H&=\frac{1}{2}\sum_{m,n}E_n\sum_{\sigma}\biggl\{ \\
	&\times\gamma_m^\dagger\gamma_n\biggl[\sum_i\int dy_i\left[u_{im\sigma}^*(y_i)u_{in\sigma}(y_i)+v^*_{im\sigma}(y_i)v_{in\sigma}(y_i)\right]
+\int d\x\left[u^*_{sm\sigma}(\x)u_{sn\sigma}(\x)+v^*_{sm\sigma}(\x)v_{sn\sigma}(\x)\right]\biggr] \\
	&-\gamma_m\gamma_n^\dagger\biggl[\sum_i\int dy_i\left[u_{im\sigma}(y_i)u_{in\sigma}^*(y_i)+v_{im\sigma}(y_i)v^*_{in\sigma}(y_i)\right]+\int d\x\left[u_{sm\sigma}(\x)u^*_{sn\sigma}(\x)+v_{sm\sigma}(\x)v^*_{sn\sigma}(\x)\right]\biggr] \\
	&+\gamma_m\gamma_n\biggl[\sum_i\int dy_i\left[u_{im\sigma}(y_i)v_{in\sigma}(y_i)+v_{im\sigma}(y_i)u_{in\sigma}(y_i)\right]+\int d\x\left[u_{sm\sigma}(\x)v_{sn\sigma}(\x)+v_{sm\sigma}(\x)u_{sn\sigma}(\x)\right]\biggr] \\
	&-\gamma_m^\dagger\gamma_n^\dagger\biggl[\sum_i\left[u_{im\sigma}^*(y_i)v^*_{in\sigma}(y_i)+v^*_{im\sigma}(y_i)u^*_{in\sigma}(y_i)\right]+\int d\x\left[u^*_{sm\sigma}(\x)v^*_{sn\sigma}(\x)+v^*_{sm\sigma}(\x)u^*_{sn\sigma}(\x)\right]\biggr] \biggr\}.
\end{aligned}
\end{equation}
Making use of Eqs.~(\ref{relationssupp}) to evaluate $\sum_{i,\sigma}\int dy_i\int d\x$, we find
\begin{equation}
H=\sum_nE_n\gamma_n^\dagger\gamma_n.
\end{equation}
Hence, our Bogoliubov transformation indeed diagonalizes the Hamiltonian.

We will now specify an explicit geometry and model for the tunneling Hamiltonian to coincide with what is considered in the main text. We assume that the system is infinite in the $y$-direction; owing to the translational invariance, the momentum $k_y$ is then a good quantum number. We assume that tunneling preserves the spin and is local in space,
\begin{equation} \label{tunnelingsupp}
T_{\sigma\sigma'}^i(\x,y_i)=-t_i\delta_{\sigma\sigma'}\delta(y-y_i)\delta(x-x_i),
\end{equation}
where $x_i$ denotes the position of wire $i$. Such a tunneling term also preserves the momentum $k_y$. Fourier transforming the BdG equations to momentum space and substituting Eq.~\eqref{tunnelingsupp}, we arrive at the form quoted in the main text:
\begin{equation} \label{BdGsupp}
\begin{gathered}
{\mathcal{H}}_i(k_y)u_i(k_y)-t_iu_s(k_y,x_i)=Eu_i(k_y,x_i), \\
{\mathcal{H}}_i^T(k_y)v_i(k_y)-t_iv_s(k_y,x_i)=-Ev_i(k_y,x_i),  \\
H_0(k_y,x)u_s(k_y,x)+i\sigma_y{\Delta}v_s(-k_y,x)-\sum_it_i\delta(x-x_i)u_i(k_y)=Eu_s(k_y,x), \\
H_0(k_y,x)v_s(k_y,x)+i\sigma_y{\Delta}^* u_s(-k_y,x)-\sum_it_i\delta(x-x_i)v_i(k_y)=-Ev_s(k_y,x), 
\end{gathered}
\end{equation}
where we introduce the electron and hole spinors $u_j=[u_{j\uparrow},u_{j\downarrow}]^T$ and $v_j=[v_{j\uparrow},v_{j\downarrow}]^T$ (and similarly for $u_s$ and $v_s$) and suppress explicit reference to quantum number $n$.

\section{Solving BdG Equations to Determine Excitation Gap} \label{Sec2supp}
To determine the excitation spectrum of Hamiltonian (\ref{Hamiltoniansupp}), we must solve Eqs.~(\ref{BdGsupp}) for $E(k_y)$. We can solve algebraically for the wave functions of the wires to yield
\begin{equation} \label{BdGsolsupp}
\begin{gathered}
u_i(k_y)=\bigl[{\mathcal{H}}_i(k_y)-E\bigr]^{-1}t_iu_s(k_y,x_i), \\
v_i(k_y)=\bigl[{\mathcal{H}}_i^T(k_y)+E\bigr]^{-1}t_iv_s(k_y,x_i).
\end{gathered}
\end{equation}
Substituting Eqs.~(\ref{BdGsolsupp}) into Eqs.~(\ref{BdGsupp}), we obtain a set of coupled differential equations describing the superconducting wave functions,
\begin{equation} \label{BdGSCsupp}
\begin{gathered}
\biggl(H_0-E-\bigl[{\mathcal{H}}_L(k_y)-E\bigr]^{-1}t_L^2\delta(x-x_L)-\bigl[\hat{\mathcal{H}}_R(k_y)-E\bigr]^{-1}t_R^2\delta(x-x_R)\biggr)u_s(k_y,x)+\hat\Delta v_s(-k_y,x)=0, \\
\biggl(-H_0-E+\bigl[{\mathcal{H}}_L^T(-k_y)+E\bigr]^{-1}t_L^2\delta(x-x_L)+\bigl[\hat{\mathcal{H}}_R^T(-k_y)+E\bigr]^{-1}t_R^2\delta(x-x_R)\biggr)v_s(-k_y,x)+\hat\Delta^\dagger u_s(k_y,x)=0.
\end{gathered}
\end{equation}
The nanowires induce delta-function terms in the BdG equations that describe the superconducting wave functions, meaning that the presence of the wires simply introduce boundary conditions to the BdG equations. These boundary conditions can be found by directly integrating Eqs.~(\ref{BdGSCsupp}):
\begin{equation} \label{BCsupp}
\begin{gathered}
\partial_xu_s(k_y,x_L)=-[{\mathcal{H}}_L(k_y)-E]^{-1}2m_st_L^2u_s(k_y,x_L), \\
\partial_xv_s(-k_y,x_L)=-[{\mathcal{H}}_L^T(-k_y)+E]^{-1}2m_st_L^2v_s(-k_y,x_L), \\
\partial_xu_s(k_y,x_R)=[{\mathcal{H}}_R(k_y)-E]^{-1}2m_st_R^2u_s(k_y,x_R), \\
\partial_xv_s(-k_y,x_R)=[{\mathcal{H}}_R^T(-k_y)+E]^{-1}2m_st_R^2v_s(-k_y,x_R).
\end{gathered}
\end{equation}
In addition to Eqs.~\eqref{BCsupp}, we also impose continuity of the superconducting wave function at $x=x_i$ and vanishing boundary conditions at the ends of the superconductor, $\psi_s(k_y,0)=\psi_s(k_y,d)=0$.

We now specialize to the case where the nanowire Hamiltonians are given by $\hat{\mathcal{H}}_L(k_y)=\hat{\mathcal{H}}_R(k_y)=\xi_k$, where $\xi_k=k_y^2/2m_n-\mu_n$ ($m_n$ and $\mu_n$ are the effective mass and Fermi energy of the nanowires, respectively), which is the case considered in the main text. Because the nanowire Hamiltonian is spin-degenerate, we can remove the trivial spin sector to reduce the dimensionality of our BdG equation; i.e., we can look only at the equations coupling $u_{s\uparrow}$ and $v_{s\downarrow}$. In the superconductor, we must solve the BdG equations in the left $l$ ($0<x<x_L$), middle $m$ ($x_L<x<x_R$), and right $r$ ($x>x_R$) regions. We find the solutions
\begin{equation}
\begin{gathered}
\psi_l(k_y,x)=c_1\begin{pmatrix} u_0 \\ v_0 \end{pmatrix}\sin(p_+x)+c_2\begin{pmatrix} v_0 \\ u_0 \end{pmatrix}\sin(p_-x) \\
\psi_m(k_y,x)=c_3\begin{pmatrix} u_0 \\ v_0 \end{pmatrix}e^{ip_+x}+c_4\begin{pmatrix} u_0 \\ v_0 \end{pmatrix}e^{-ip_+x}+c_5\begin{pmatrix} v_0 \\ u_0 \end{pmatrix}e^{ip_-x}+c_6\begin{pmatrix} v_0 \\ u_0 \end{pmatrix}e^{-ip_-x}, \\
\psi_r(k_y,x)=c_7\begin{pmatrix} u_0 \\ v_0 \end{pmatrix}\sin[p_+(d-x)]+c_8\begin{pmatrix} v_0 \\ u_0 \end{pmatrix}\sin[p_-(d-x)],
\end{gathered}
\end{equation}
where $\psi(k_y,x)=[u_{s\uparrow}(k_y,x),v_{s\downarrow}(-k_y,x)]^T$ is a spinor wave function, $p_\pm^2=k_F^2-k_y^2\pm 2im_s\Omega$ ($k_F=\sqrt{2m_s\mu_s}$ is the Fermi momentum of the superconductor), $\Omega^2=\Delta^2-E^2$, and $u_0^2(v_0^2)=(1\pm i\Omega/E)/2$ are the usual BCS coherence factors. For our choice of nanowire Hamiltonian, the boundary conditions Eqs.~(\ref{BCsupp}) take the form quoted in the main text. In solving the boundary conditions, we take $k_y\lesssim k_{Fn}\ll k_F$; i.e., we assume that the density of the nanowires is much smaller than the density of the superconductor and because $k_y$ is conserved in tunneling typical values should not exceed the Fermi momentum of the wires. We additionally expand $p_\pm$ in the limit $\Delta\ll\mu_s$. Making use of these two approximations, we expand $p_\pm=k_F$ outside of the exponentials and $p_\pm=k_F\pm i\Omega/v_F$ inside the exponentials ($v_F=k_F/m_s$). We additionally simplify the problem by taking a symmetric nanowire configuration, $x_L=x_w$ and $x_R=d-x_w$, and assuming that the nanowires are located close to the boundaries of the superconductor, $x_w\ll d$. The latter assumption allows us to replace $\Omega x_w/v_F\to0$. With these simplifications, the boundary conditions can be expressed in matrix form as $Mc=0$, where $c$ is a vector of unknown coefficients and
\begin{equation} \label{BCmat}
\begin{aligned}
\small
M=\left(\begin{array}{cccc}
	-u_0\left[\cos(k_Fx_w)+\frac{2\gamma_L}{E-\xi_k}\sin(k_Fx_w)\right] & -v_0\left[\cos(k_Fx_w)+\frac{2\gamma_L}{E-\xi_k}\sin(k_Fx_w)\right] & iu_0e^{ik_Fx_w} & -iu_0e^{-ik_Fx_w} \\
	-v_0\left[\cos(k_Fx_w)-\frac{2\gamma_L}{E+\xi_k}\sin(k_Fx_w)\right] & -u_0\left[\cos(k_Fx_w)-\frac{2\gamma_L}{E+\xi_k}\sin(k_Fx_w)\right] & iv_0e^{ik_Fx_w} & -iv_0e^{-ik_Fx_w} \\
	0 & 0 & iu_0e^{ik_F(d-x_w)}e^{-\chi} & -iu_0e^{-ik_F(d-x_w)}e^{\chi} \\
	0 & 0 & iv_0e^{ik_F(d-x_w)}e^{-\chi} & -iv_0e^{-ik_F(d-x_w)}e^{\chi} \\
	-u_0\sin(k_Fx_w) & -v_0\sin(k_Fx_w) & u_0e^{ik_Fx_w} & u_0e^{-ik_Fx_w} \\
	-v_0\sin(k_Fx_w) & -u_0\sin(k_Fx_w) & v_0e^{ik_Fx_w} & v_0e^{-ik_Fx_w} \\
	0 & 0 & u_0e^{ik_F(d-x_w)}e^{-\chi} & u_0e^{-ik_F(d-x_w)}e^\chi \\
	0 & 0 & v_0e^{ik_F(d-x_w)}e^{-\chi} & v_0e^{-ik_F(d-x_w)}e^\chi
	\end{array}\right.\cdots \\
	\left.\begin{array}{cccc}
	iv_0e^{ik_Fx_w} & -iv_0e^{-ik_Fx_w} & 0 & 0 \\
	iu_0e^{ik_Fx_w} & -iu_0e^{-ik_Fx_w} & 0 & 0 \\
	iv_0e^{ik_F(d-x_w)}e^\chi & -iv_0e^{-ik_F(d-x_w)}e^{-\chi} & u_0\left[\cos(k_Fx_w)+\frac{2\gamma_R}{E-\xi_k}\sin(k_Fx_w)\right] & v_0\left[\cos(k_Fx_w)+\frac{2\gamma_R}{E-\xi_k}\sin(k_Fx_w)\right] \\
	iu_0e^{ik_F(d-x_w)}e^{\chi} & -iu_0e^{-ik_F(d-x_w)}e^{-\chi} & v_0\left[\cos(k_Fx_w)-\frac{2\gamma_R}{E+\xi_k}\sin(k_Fx_w)\right] & u_0\left[\cos(k_Fx_w)-\frac{2\gamma_R}{E+\xi_k}\sin(k_Fx_w)\right] \\
	v_0e^{ik_Fx_w} & v_0e^{-ik_Fx_w} & 0 & 0 \\
	u_0e^{ik_Fx_w} & u_0e^{-ik_Fx_w} & 0 & 0 \\
	v_0e^{ik_F(d-x_w)}e^\chi & v_0e^{-ik_F(d-x_w)}e^{-\chi} & -u_0\sin(k_Fx_w) & -v_0\sin(k_Fx_w) \\
	u_0e^{ik_F(d-x_w)}e^\chi & u_0 e^{-ik_F(d-x_w)}e^{-\chi} & -v_0\sin(k_Fx_w) & -u_0\sin(k_Fx_w)
	\end{array}\right),
\end{aligned}
\end{equation}
where we introduce the tunneling energy scale $\gamma_i=t_i^2/v_F$ and the shorthand $\chi=\Omega d/v_F$. 

In order to proceed analytically, we assume that the nanowires are only weakly coupled to the superconductor ($\gamma_i\ll\Delta$). Then the relevant pairing energies in the nanowires should be much smaller than the superconducting gap and we can look only at low energies $E\ll\Delta$. Making this assumption, we replace $\chi=d/\xi_s$ (with $\xi_s=v_F/\Delta$ the coherence length of the superconductor) and $u_0^2(v_0^2)=\pm i\Delta/(2E)$. The condition for the solvability of Eq.~(\ref{BCmat}) is then given by
\begin{equation} \label{solvability1}
\begin{aligned}
0&=E^4-\frac{2E^2}{\cosh(2\chi)-\cos(2k_Fd)}\biggl\{\xi_k^2[\cosh(2\chi)-\cos(2k_Fd)]-\xi_k(\gamma_L+\gamma_R)\bigl\{\cosh(2\chi)\sin(2k_Fx_w) \\
	&-2\cos[k_F(2d-x_w)]\sin(k_Fx_w)\bigr\}+2\sin^2(k_Fx_w)\bigl[(\gamma_L^2+\gamma_R^2)\bigl\{\cosh(2\chi)-\cos[2k_F(d-x_w)]\bigr\}+4\gamma_L\gamma_R\sin^2(k_Fx_w)\bigr]\biggr\} \\
	&+\frac{1}{\cosh(2\chi)-\cos(2k_Fd)}\biggl\{\xi_k^4[\cosh(2\chi)-\cos(2k_Fd)]-2\xi_k^3(\gamma_L+\gamma_R)\bigl\{\cosh(2\chi)\sin(2k_Fx_w)-2\cos[k_F(2d-x_w)]\sin(k_Fx_w)\bigr\} \\
	&+4\xi_k^2\sin^2(k_Fx_w)\{[(\gamma_L+\gamma_R)^2+2\gamma_L\gamma_R\cos(2k_Fx_w)]\cosh(2\chi)-(\gamma_L^2+4\gamma_L\gamma_R+\gamma_R^2)\cos[2k_F(d-x_w)]\bigr\} \\
	&-16\xi_k\gamma_L\gamma_R(\gamma_L+\gamma_R)\sin^3(k_Fx_w)\bigl\{\cosh(2\chi)\cos(k_Fx_w)-\cos[k_F(2d-3x_w)]\bigr\} \\
	&+16\gamma_L^2\gamma_R^2\sin^4(k_Fx_w)\bigl\{\cosh(2\chi)-\cos[2k_F(d-2x_w)]\bigr\}\biggr\}.
\end{aligned}
\end{equation}
Equation~(\ref{solvability1}) is nothing more than a quadratic equation for $E^2$ that can be solved exactly. However, it is still difficult to calculate the gap in the excitation spectrum for an arbitrary superconductor width $d$ and asymmetric tunneling $\gamma_L\neq\gamma_R$. We therefore further constrain our system by specifying several different limits that we can treat completely analytically.

\subsection{Symmetric Tunneling ($\gamma_L=\gamma_R$)}
When $\gamma_L=\gamma_R$, we solve Eq.~(\ref{solvability1}) to yield the spectrum
\begin{equation} \label{spectrumsymm}
\begin{aligned}
E^2_\pm(\xi_k)&=\frac{1}{\cosh(2\chi)-\cos (2k_Fd)}\biggl\{\xi_k^2[\cosh(2\chi)-\cos (2k_Fd)]-2\xi_k\gamma\bigl\{\cosh(2\chi)\sin(2k_Fx_w)-2\cos[k_F(2d-x_w)]\sin(k_Fx_w)\bigr\} \\
	&+4\gamma^2\sin^2(k_Fx_w)\bigl\{\cosh(2\chi)-\cos[2k_F(d-x_w)]+2\sin^2(k_Fx_w)\bigr\}		
		\pm8\gamma\cosh\chi\sin^2(k_Fx_w) \\
	&\times\sqrt{\{2\gamma\sin[k_F(d-x_w)]\sin(k_Fx_w)-\xi_k\sin (2k_Fd)\}^2}\biggr].
\end{aligned}
\end{equation}
Differentiating the lower branch of the spectrum $E_-^2$, we find that it has two local minima near
\begin{equation}
\xi_\pm=\frac{\gamma\sin (k_Fx_w)}{\cosh\chi\pm\cos (k_Fd)}\{\pm\cosh\chi\cos(k_Fx_w)+\cos[k_F(d-x_w)]\}.
\end{equation}
Substituting both $\xi_+$ and $\xi_-$ into Eq.~(\ref{spectrumsymm}) to determine the global minimum of $E_-(\xi_k)$, we obtain a gap of the form
\begin{equation}
E_g=\frac{2\gamma\sinh\chi\sin^2(k_Fx_w)}{\cosh\chi+|\cos (k_Fd)|}.
\end{equation}
The gap oscillates on the scale of the Fermi wavelength and approaches $E_g=2\gamma\sin^2(k_Fx_w)$ when the superconductor is very wide ($\chi\gg1$). When the superconductor is very narrow ($\chi\ll1$), we expand to give
\begin{equation}
E_g=\frac{2\gamma\chi\sin^2(k_Fx_w)}{1+|\cos (k_Fd)|}.
\end{equation}
In this limit, the gap oscillates between its maximum value $E_g^\text{max}=2\gamma\chi\sin^2(k_Fx_w)$ and its minimum value $E_g^\text{min}=\gamma\chi\sin^2(k_Fx_w)$.

\subsection{One Wire Decoupled (Direct Pairing Only)}
Suppose that we decouple one of the nanowires by setting $\gamma_L=0$ (and we also write $\gamma_R=\gamma$). In this case, we completely suppress crossed Andreev pairing and can investigate the gap opened in the presence of direct pairing only. Setting $\gamma_L=0$, the solvability condition of Eq.~(\ref{solvability1}) simplifies to
\begin{equation}
\begin{aligned}
0&=(E^2-\xi_k^2)\biggl\{(E^2-\xi_k^2)[\cosh(2\chi)-\cos(2k_Fd)]+4\gamma\xi_k\sin(k_Fx_w)\{\cosh(2\chi)\cos(k_Fx_w)-\cos[k_F(2d-x_w)]\} \\
	&-4\gamma^2\sin^2(k_Fx_w)\{\cosh(2\chi)-\cos[2k_F(d-x_w)]\}\biggr\}.
\end{aligned}
\end{equation}
The first factor $(E^2-\xi_k^2)$ describes a trivial gapless spectrum and simply represents the decoupling of the left nanowire. The spectrum of the superconductor/right-nanowire system is described by the terms enclosed by braces. Solving for the spectrum, we find
\begin{equation} \label{speconewire}
E^2=\xi_k^2-\frac{4\gamma\sin(k_Fx_w)}{\cosh(2\chi)-\cos(2k_Fd)}\biggl[\xi_k\{\cosh(2\chi)\cos(k_Fx_w)-\cos[k_F(2d-x_w)]\}-\gamma\sin(k_Fx_w)\{\cosh(2\chi)-\cos[2k_F(d-x_w)]\}\biggr].
\end{equation}
Differentiating with respect to $\xi_k$, we find that the spectrum is minimized for
\begin{equation} \label{suppminonewire}
\xi_k=-\frac{\gamma}{\cosh(2\chi)-\cos (2k_Fd)}\bigl\{\sin(2k_Fd)-\cosh(2\chi)\sin(2k_Fx_w)-\sin[2k_F(d-x_w)]\bigr\}.
\end{equation}
We substitute this value of $\xi_k$ into Eq.~(\ref{speconewire}) to find a direct gap given by
\begin{equation} \label{directgapsupp}
E_g^D=\frac{2\gamma\sinh(2\chi)\sin^2(k_Fx_w)}{\cosh(2\chi)-\cos (2k_Fd)}.
\end{equation}
When the superconductor is very narrow ($\chi\ll1$), the gap is sharply peaked near $k_Fd=n\pi$ and is minimized near $k_Fd=\pi(n+1/2)$ for $n\in\mathbb{Z}$,
\begin{equation} \label{Eg1w}
E_g^D=\frac{4\gamma\chi\sin^2(k_Fx_w)}{1-\cos (2k_Fd)+2\chi^2}.
\end{equation}
Expanding near the extrema, we find that the gap oscillates between the maximum value $E_g^{D,\text{max}}=2\gamma\sin^2(k_Fx_w)/\chi$ and the minimum value $E_g^{D,\text{min}}=2\gamma\chi\sin^2(k_Fx_w)$. When the superconductor is very wide ($\chi\gg1$), the gap approaches $E_g^D=2\gamma\sin^2(k_Fx_w)$, as it should.

\subsection{Quantum Hall Regime (Crossed Andreev Pairing Only)} \label{QuantumHallSupp}
In order to isolate crossed Andreev pairing, we imagine that our nanowires are spin-polarized. This corresponds to removing the spin sum in both Eqs.~(\ref{HNWsupp}) and (\ref{Htsupp}); explicitly, we assume that the left nanowire contains only spin-up states and the right wire contains only spin-down states. In the BdG equations [Eqs.~\eqref{BdGsupp}], we set $t_R=0$ in the equations describing spin-up states and we set $t_L=0$ in the equations describing spin-down states. In the boundary condition matrix Eq.~\eqref{BCmat}, this amounts to setting $\gamma_L=0$ in the second column and $\gamma_R=0$ in the seventh column. 

Assuming that the tunneling strengths are equation, $\gamma_L=\gamma_R$, we find a solvability condition given by
\begin{equation}
\small
\begin{aligned}
0&=(E^2-\xi_k^2)\biggl\{(E^2-\xi_k^2)[\cosh(2\chi)-\cos(2k_Fd)]^2+4\gamma\xi_k\sin(k_Fx_w)\{\cosh(2\chi)\cos(k_Fx_w)-\cos[k_F(2d-x_w)]\} \\
	&\times[\cosh(2\chi)-\cos(2k_Fd)]-\gamma^2\bigl[\{\cosh(2\chi)\sin(2k_Fx_w)-2\cos[k_F(2d-x_w)]\sin(k_Fx_w)\}^2+16\sinh^2\chi\cos^2(k_Fd)\sin^4(k_Fx_w)\bigr]\biggr\}.
\end{aligned}
\end{equation}
Similarly to the previous section, there is an overall factor $(E^2-\xi_k^2)$ which represents the spectrum of the two decoupled spin channels. Solving for the spectrum of the coupled channels, we have
\begin{equation}
\begin{aligned}
E^2&=\xi_k^2-\frac{4\gamma\xi_k\sin(k_Fx_w)}{\cosh(2\chi)-\cos(2k_Fd)}\{\cosh(2\chi)\cos(k_Fx_w)-\cos[k_F(2d-x_w)]\}+\frac{\gamma^2}{[\cosh(2\chi)-\cos(2k_Fd)]^2} \\
&\times\bigl[\{\cosh(2\chi)\sin(2k_Fx_w)-2\cos[k_F(2d-x_w)]\sin(k_Fx_w)\}^2+16\sinh^2\chi\cos^2(k_Fd)\sin^4(k_Fx_w)\bigr].
\end{aligned}
\end{equation}
The spectrum is once again minimized by choosing
\begin{equation}
\xi_k=-\frac{\gamma}{\cosh(2\chi)-\cos (2k_Fd)}\bigl\{\sin(2k_Fd)-\cosh(2\chi)\sin(2k_Fx_w)-\sin[2k_F(d-x_w)]\bigr\}.
\end{equation}
Substituting this value of $\xi_k$, we find a crossed Andreev gap given by
\begin{equation} \label{CAgapsupp}
E_g^C=\frac{4\gamma|\cos (k_Fd)|\sinh\chi\sin^2(k_Fx_w)}{\cosh(2\chi)-\cos (2k_Fd)}.
\end{equation}
When the superconductor is very narrow ($\chi\ll1$), we expand the gap to give
\begin{equation}
E_g^C=\frac{4\gamma\chi|\cos (k_Fd)|\sin^2(k_Fx_w)}{1-\cos (2k_Fd)+2\chi^2}.
\end{equation}
The gap is maximized when $k_Fd=n\pi$, where it takes the value $E_g^C=2\gamma|\cos (k_Fd)|\sin^2(k_Fx_w)/\chi$. The crossed Andreev gap vanishes when $k_Fd=\pi(n+1/2)$. When the superconductor is very wide ($\chi\gg1$), the gap decays exponentially with $d$ on the scale of $\xi_s$ and oscillates on the scale of $1/k_F$, $E_g^C=4\gamma|\cos (k_Fd)|\sin^2(k_Fx_w)e^{-\chi}$.

\section{Mapping to Effective Pairing Model} \label{Sec3supp}
In this section, we reinterpret our exact solution for the excitation spectrum of the double-nanowire system (displayed in Fig.~2 of the main text) in terms of an effective pairing model similar to those used in Refs.~[\onlinecite{Klinovaja:2014,Klinovaja:2014b}]. The Hamiltonian of the nanowires is given by
\begin{equation}
H_{NW}=\sum_\sigma\sum_{i=L,R}\int\frac{dk_y}{2\pi}\,\psi_{i\sigma}^\dagger(k_y)\xi_k\psi_{i\sigma}(k_y).
\end{equation}
Direct pairing is incorporated through the inclusion of an intrinsic superconducting term in the Hamiltonian,
\begin{equation}
H_d^i=\Delta_i\int\frac{dk_y}{2\pi}\left[\psi^\dagger_{i\uparrow}(k_y)\psi^\dagger_{i\downarrow}(-k_y)+H.c.\right].
\end{equation}
Similarly, crossed Andreev pairing is incorporated through a term
\begin{equation}
H_c=\Delta_c\sum_{i=L,R}\int\frac{dk_y}{2\pi}\left[\psi_{i\uparrow}^\dagger(k_y)\psi_{\bar{i}\downarrow}^\dagger(-k_y)+H.c.\right],
\end{equation}
where $\bar{i}$ denotes the opposite wire as $i$. Finally, single-particle intrawire and interwire couplings can be incorporated through the inclusion of the terms
\begin{subequations}
\begin{gather}
H_{\delta\mu}^i=-\delta\mu_i\sum_\sigma\int\frac{dk_y}{2\pi}\left[\psi_{i\sigma}^\dagger(k_y)\psi_{i\sigma}(k_y)+H.c.\right], \\
H_\Gamma=-\Gamma\sum_\sigma\sum_{i=L,R}\int\frac{dk_y}{2\pi}\left[\psi_{i\sigma}^\dagger(k_y)\psi_{\bar{i}\sigma}(k_y)+H.c.\right].
\end{gather}
\end{subequations}
The effective parameters $\Delta_i$ and $\delta\mu_i$ are proportional to $t_i^2$ while $\Delta_c$ and $\Gamma$ are proportional to $t_Lt_R$, and all four parameters are unknown functions of the superconductor width. The total Hamiltonian can be expressed as
\begin{equation} \label{effectivemodelH}
H=\int\frac{dk_y}{2\pi}\,\Psi^\dagger(k_y)\left(\begin{array}{cccc}
	\xi_k-\delta\mu_L & \Delta_L & -\Gamma & \Delta_c \\
	\Delta_L & -\xi_k+\delta\mu_L & \Delta_c & \Gamma \\
	-\Gamma & \Delta_c & \xi_k-\delta\mu_R & \Delta_R \\
	\Delta_c & \Gamma & \Delta_R & -\xi_k+\delta\mu_R \end{array}\right)\Psi(k_y),
\end{equation}
where we define the spinor $\Psi(k_y)=[\psi_{L\uparrow}(k_y),\psi_{L\downarrow}^\dagger(-k_y),\psi_{R\uparrow}(k_y),\psi_{R\downarrow}^\dagger(-k_y)]^T$. The spectrum of the effective model can be found by solving the following equation:
\begin{equation} \label{effectivemodeleq}
\begin{aligned}
0&=E^4-E^2\bigl[2\Gamma^2+2\Delta_c^2+\Delta_L^2+\Delta_R^2+\delta\mu_L^2+\delta\mu_R^2+2\xi_k^2-2(\delta\mu_L+\delta\mu_R)\xi_k\bigr] \\
	&+\Gamma^4+\Delta_c^4+[\Delta_L^2+(\xi_k-\delta\mu_L)^2][\Delta_R^2+(\xi_k-\delta\mu_R)^2]-4\Gamma\Delta_c[\Delta_r\delta\mu_L+\Delta_L\delta\mu_R-(\Delta_L+\Delta_R)\xi_k] \\
		&-2\Delta_c^2[\Delta_L\Delta_R-(\xi_k-\delta\mu_L)(\xi_k-\delta\mu_R)]+2\Gamma^2[\Delta_c^2+\Delta_L\Delta_R-(\xi_k-\delta\mu_L)(\xi_k-\delta\mu_R)].
\end{aligned}
\end{equation}
We now look to directly map Eq.~(\ref{effectivemodeleq}) onto Eq.~(\ref{solvability1}). 

To determine the correct mapping, let us first consider the case where wire $\bar i$ is decoupled from the superconductor. In the effective model, this corresponds to setting $\Gamma=\Delta_c=\Delta_{\bar i}=\delta\mu_{\bar i}=0$; in the BdG solution, we set $\gamma_{\bar i}=0$. The spectrum of the effective model in this case is given by
\begin{equation}
E^2=\Delta_i^2+(\xi_k-\delta\mu_i)^2.
\end{equation}
The band minimum is given by $\delta\mu_i$, while the gap is given by $\Delta_i$. Comparing with Eqs.~(\ref{suppminonewire}) and (\ref{directgapsupp}), we find
\begin{subequations}
\begin{gather}
\delta\mu_i=-\frac{\gamma_i}{\cosh(2\chi)-\cos (2k_Fd)}\bigl\{\sin(2k_Fd)-\cosh(2\chi)\sin(2k_Fx_w)-\sin[2k_F(d-x_w)]\bigr\}, \label{deltamueff} \\
\Delta_i=\frac{2\gamma_i\sinh(2\chi)\sin^2(k_Fx_w)}{\cosh(2\chi)-\cos (2k_Fd)}. \label{deltaDeff}
\end{gather}
\end{subequations}

Now let us restore the second wire. Comparing the $E^2$ terms of Eqs.~(\ref{effectivemodeleq}) and (\ref{solvability1}) given the expressions for $\delta\mu_i$ and $\Delta_i$ in Eqs.~(\ref{deltamueff}) and (\ref{deltaDeff}), we see that the remaining effective parameters must satisfy
\begin{equation}
\Delta_c^2+\Gamma^2=\frac{8\gamma_L\gamma_R\sin^4(k_Fx_w)}{\cosh(2\chi)-\cos (2k_Fd)}, \label{suppconstraint1}
\end{equation}
We also compare the $E^0$ terms of Eqs.~(\ref{effectivemodeleq}) and (\ref{solvability1}) to find the additional constraints
\begin{subequations}
\begin{gather}
\Delta_c^2-\Gamma^2=\frac{8\gamma_L\gamma_R[\cosh(2\chi)\cos (2k_Fd)-1]\sin^4(k_Fx_w)}{[\cosh(2\chi)-\cos (2k_Fd)]^2}, \label{suppconstraint2} \\
\Delta_c\Gamma=-\frac{4\gamma_L\gamma_R\sinh(2\chi)\sin (2k_Fd)\sin^4(k_Fx_w)}{[\cosh(2\chi)-\cos (2k_Fd)]^2}. \label{suppconstraint3}
\end{gather}
\end{subequations}
Equations~(\ref{suppconstraint1}) and (\ref{suppconstraint2}) determine the magnitudes of $\Delta_c$ and $\Gamma$, while Eq.~(\ref{suppconstraint3}) determines the relative sign. Solving, we obtain
\begin{subequations}
\begin{gather}
\Gamma=-\frac{4\sqrt{\gamma_L\gamma_R}\cosh\chi\sin (k_Fd)\sin^2(k_Fx_w)}{\cosh(2\chi)-\cos (2k_Fd)}, \label{Gammaeff} \\
\Delta_c=\frac{4\sqrt{\gamma_L\gamma_R}\sinh\chi\cos (k_Fd)\sin^2(k_Fx_w)}{\cosh(2\chi)-\cos (2k_Fd)}. \label{deltaCeff}
\end{gather}
\end{subequations}
Comparing with the direct and crossed Andreev excitation gaps that we found by solving the BdG equations in the limit $\gamma_L=\gamma_R$ [Eqs.~(\ref{directgapsupp}) and (\ref{CAgapsupp}), respectively], we see that $E_g^D=\Delta_i$ and $E_g^C=|\Delta_c|$.

We have thus mapped an effective pairing model onto our exact solution for the excitation spectrum of the double-nanowire system. Interpretation of the spectrum displayed in Fig. 2 of the main text is most easily seen if we take $\Delta_L=\Delta_R=\Delta_d$ and $\delta\mu_L=\delta\mu_R$. In this case, the spectrum of the effective model is given by
\begin{equation}
E_\pm^2=(\Delta_d\pm\Delta_c)^2+(\xi_k-\delta\mu\mp\Gamma)^2.
\end{equation}
We see that tunneling shifts the band minima away from $\xi_k=0$ to $\xi_k=\delta\mu\pm\Gamma$ while inducing an excitation gap $E_g=\Delta_d-|\Delta_c|$. Therefore, when the superconductor is very wide ($\chi\gg1$), tunneling induces only direct pairing in the nanowires. As the wires are brought closer together, crossed Andreev pairing reduces the size of the gap and single-particle couplings shift the effective chemical potential of each band $\mu\to\mu+\delta\mu\pm\Gamma$.

\section{Tight-binding calculations}
In the main text, we presented a numerical tight-binding calculation that supported our analytical calculations on the system of two nanowires separated by a finite-sized $s$-wave superconductor. Here we present the details of this calculation as well as some additional numerical calculations that probe limits that are not easily treated analytically within our BdG solution. Then, once we gain some insights from the numerics, we will utilize the effective pairing model derived in Sec.~\ref{Sec3supp} to make some analytical statements about the numerical results where possible.

\subsection{Tight-Binding Hamiltonian}

We construct the numerical tight-binding model for proximity-induced superconductivity [\onlinecite{Rainis:2013,Klinovaja:2015}] in the geometry shown in Fig.~1 of the main text. Because the system is assumed infinite in the $y$-direction, the Hamiltonian takes a block-diagonal form in momentum $k_y$, 
\begin{equation}
H=\sum_{k_y}H_{k_y}.
\end{equation}
The size of the system in the $x$ direction is $(N+2) a_x$, where $a_{x,y}$ are the lattice constants and $(N+2)$ is the number of sites (the left wire corresponds to site 1 while the right wire to site $N+2$). The kinetic part of the Hamiltonian is given by
\begin{equation}
\begin{aligned}
H_{0,k_y}&= -\sum_\sigma \sum_{i=1}^{N+1}(t_i c_{k_y,i+1,\sigma}^\dagger c_{k_y,i,\sigma} + H.c.)+ \sum_\sigma \sum_{i=1}^{N+2} [\mu_i- 2 t \cos (k_y a_y)] c_{k_y,i,\sigma}^\dagger c_{k_y,i,\sigma},\end{aligned}
\end{equation}
where the tunneling amplitudes are given by $t_1=t_L$, $t_{N+1}=t_R$, and $t_i=t$ ($1<i<N+1$). The Fermi energies, as measured from the bottom of the band, are given by $\mu_1=\mu_{N+2}=\mu_n$ and $\mu_i=\mu_s$ ($1<i<N+2$). The Hamiltonian of the superconductor also contains a pairing term, $H=H_0+H_{sc}$, written as
\begin{equation}
H_{sc,k_y} = \sum_{i=2}^{N+1}(\Delta c_{k_y ,i ,\uparrow}^\dagger c_{-k_y ,i ,\downarrow}^\dagger + H.c.).
\end{equation}
Following our analytical calculations, we also calculate the dependence of the proximity-induced superconducting gap in the presence of direct pairing only (achieved by setting $t_L=0$) and in the presence of crossed Andreev pairing only (achieved by setting $t_{L \downarrow}=t_{R \uparrow}=0$). The numerical results in the symmetric tunneling limit ($t_L=t_R$) is shown in Fig. 3(b) of the main text.

\begin{figure}[t!]
\includegraphics[width=\linewidth]{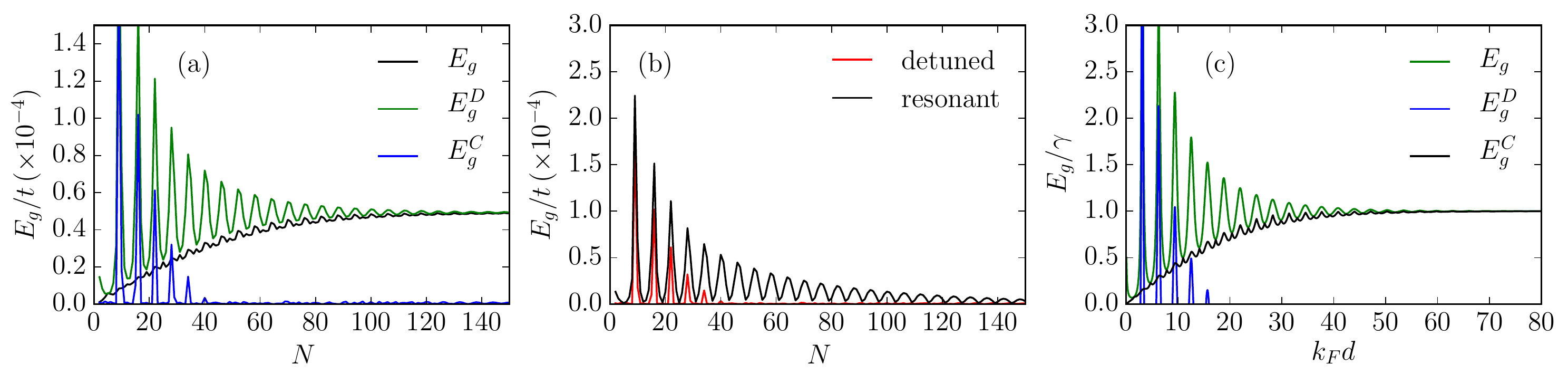}
\caption{\label{detune} Proximity-induced superconducting gaps when the wire Fermi energies are detuned, plotted as a function of superconductor width (in dimensionless units $N$ for numerics and $k_Fd$ for analytics). (a) Numerical tight-binding calculation with $\Delta=0.02t$, $\mu_s=0.3t$, $\mu_L=0.03t$, and $\mu_R=0.0301t$. Black curve corresponds to two wires ($t_L=t_R=0.01t$), green curve corresponds to single wire ($t_L=0.01t$ and $t_R=0$), and blue curve corresponds to quantum Hall regime ($t_{L\uparrow}=t_{R\downarrow}=0.01t$ and $t_{L\downarrow}=t_{R\uparrow}=0$). (b) Direct comparison of crossed Andreev gap in detuned (red) and resonant (black) [Fig.~3(b) of main text] cases. (c) Analytical results for proximity gaps, plotted with a detuning parameter $\eta=2\gamma$.}
\end{figure}

\subsection{Detuned Wire Fermi Energies}

First, we consider the case where the Fermi energies of the two nanowires are different. In this case, one would expect the direct pairing strength to be unaffected by the relative shift in Fermi energies. However, the effect of crossed Andreev pairing should be suppressed as the Fermi energies are detuned. This is precisely what we see numerically, as shown in Fig.~\ref{detune}(a-b).

Analytically, we can describe the situation where the Fermi energies are detuned using an effective Hamiltonian given by
\begin{equation} \label{Heffdetune}
\mathcal{H}_\text{eff}=\left(\begin{array}{cccc}
	\xi_k-\delta\mu & \Delta_d & -\Gamma & \Delta_c \\
	\Delta_d & -\xi_k+\delta\mu & \Delta_c & \Gamma \\
	-\Gamma & \Delta_c & \xi_k-\eta-\delta\mu & \Delta_d \\
	\Delta_c & \Gamma & \Delta_d & -\xi_k+\eta+\delta\mu
	\end{array}\right),
\end{equation}
where the effective parameters $\delta\mu$, $\Gamma$, $\Delta_d$, and $\Delta_c$ are as given in Sec.~\ref{Sec3supp}. We assume that we are in the symmetric tunneling limit ($\gamma_L=\gamma_R$) and we take the Fermi energies of the nanowires to be $\mu_L=\mu_n$ and $\mu_R=\mu_n+\eta$ (without loss of generality, we take $\eta>0$). First, we note that the direct gap is $E_g^D=\Delta_d$ regardless of which wire is decoupled. If the right wire is decoupled, then the spectrum is minimized for $\xi_k=\delta\mu$; if the left wire is decoupled, the spectrum is minimized for $\xi_k=\delta\mu+\eta$. We can calculate the crossed Andreev gap by setting $\Gamma=\Delta_d=0$ (this corresponds to the quantum Hall regime where the wires are spin-polarized with opposite spin). In this case, the spectrum is
\begin{equation}
E_\pm^2=\Delta_c^2+\frac{\eta^2}{2}-\eta(\xi_k-\delta\mu)+(\xi_k-\delta\mu)^2\pm\frac{\eta}{2}\sqrt{4\Delta_c^2+(2\xi_k-2\delta\mu-\eta)^2}.
\end{equation}
Differentiating the spectrum, we find that if $|\Delta_c|>\eta/2$, the global minimum of the spectrum is found by choosing $\xi_k=\delta\mu+\eta/2$. Choosing this value for $\xi_k$ we find a crossed Andreev gap given by
\begin{equation} \label{CAdetune}
E_g^C(d)=|\Delta_c(d)|-\frac{\eta}{2}.
\end{equation}
The detuning of the wire Fermi energies has a detrimental effect on the crossed Andreev gap, and the gap closes when the detuning becomes $\eta=2|\Delta_c|$. Physically, this is a result of the fact that crossed Andreev pairing acts to pair states in the two wires with opposite momentum $k_y$; if the Fermi energies are unequal, the opposite-momentum states in the two wires have different energies. If this energy mismatch becomes larger than the pairing $\Delta_c$ itself, then crossed Andreev pairing becomes suppressed completely. It is also possible to calculate the gap analytically in the presence of both types of pairings [keeping all effective parameters in Eq.~\eqref{Heffdetune} nonzero], but the resulting expression for the gap is very cumbersome and not particularly enlightening. Rather than give the full expression, we plot the gap in Fig.~\ref{detune}(c), showing very good agreement with the numerical results [additionally, we choose $\sin^2(k_Fx_w)=1/2$].

\begin{figure}[t!]
\includegraphics[width=.9\linewidth]{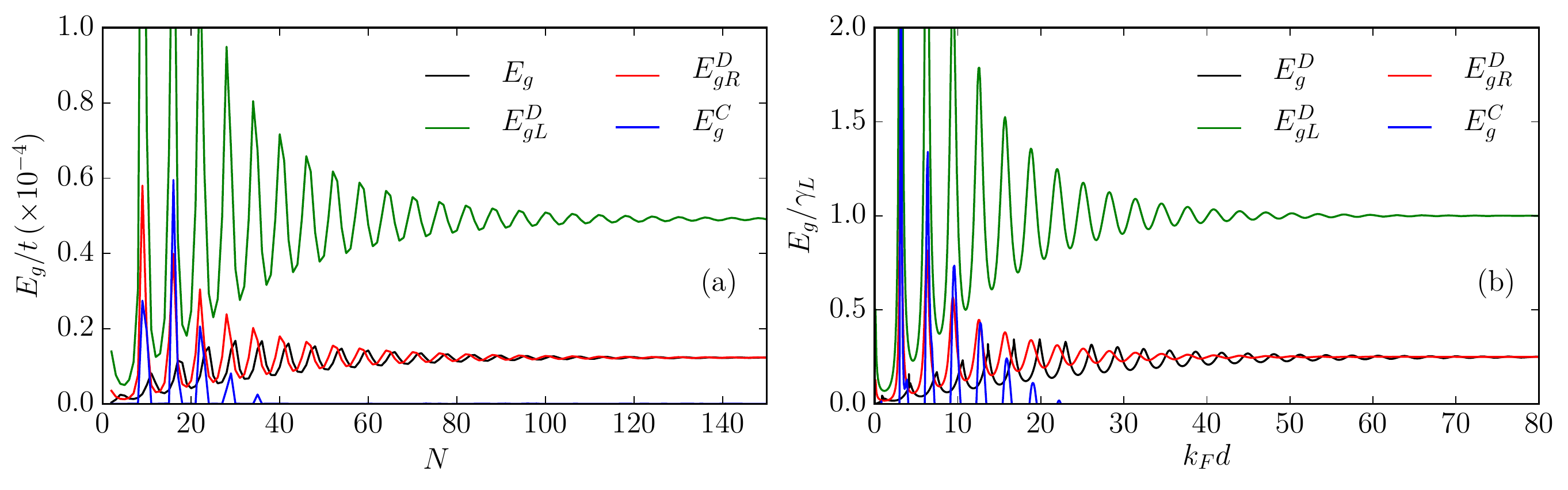}
\caption{\label{asymm} Proximity-induced superconducting gaps when the tunneling strengths are unequal ($t_L\neq t_R$), plotted as a function of superconductor width (in dimensionless units $N$ for numerics and $k_Fd$ for analytics). (a) Numerical tight-binding calculation with $\Delta=0.02t$, $\mu_s=0.3t$, $\mu_n=0.03t$, $t_L=0.01t$, and $t_R=0.005t$. (b) Analytical results for proximity gaps, plotted with $t_L=2t_R$ (equivalently, $\gamma_L=4\gamma_R$).}
\end{figure}

\subsection{Asymmetric Tunneling}

We now consider the case where tunneling is asymmetric ($t_L\neq t_R$). The numerical results are plotted in Fig.~\ref{asymm}(a). Analytically, we can make a few statements about the spectrum utilizing the effective pairing model of Sec.~\ref{Sec3supp}. We first note that the direct pairing gaps $E_{g,i}^D$ are simply given by $E_{g,i}^D=\Delta_i$. The crossed Andreev gap can be found by considering the quantum Hall regime where $\Delta_i=\Gamma=0$. In this limit, the excitation spectrum is given by
\begin{equation}
E_\pm=\frac{1}{2}\left(\delta\mu_L-\delta\mu_R\pm\sqrt{4\Delta_c^2+(\delta\mu_L+\delta\mu_R-2\xi_k)^2}\right),
\end{equation}
with $E=-E_\pm$ also corresponding to additional branches of the spectrum. We find the minimum of the spectrum at $\xi_k=(\delta\mu_L+\delta\mu_R)/2$, giving a crossed Andreev gap
\begin{equation}
E_g^C(d)=|\Delta_c(d)|-\frac{1}{2}|\delta\mu_L-\delta\mu_R|.
\end{equation}
Comparing with Eq.~\eqref{CAdetune}, we see that the asymmetric tunneling acts as an effective detuning of the Fermi energies through the asymmetric Fermi energy shifts $\delta\mu_L\neq\delta\mu_R$. Therefore, if the tunneling is made too asymmetric (when $|\delta\mu_L-\delta\mu_R|=2|\Delta_c|$), crossed Andreev pairing is completely suppressed. Of course, this detrimental effect to the crossed Andreev pairing can be compensated by appropriately adjusting the Fermi energies of the wires, $\mu_L\to\mu_n-\delta\mu_L$ and $\mu_R\to\mu_n-\delta\mu_R$. In this case, $\delta\mu_i$ drops out of the effective Hamiltonian and the crossed Andreev gap is again given by $E_g^C=|\Delta_c|$. While it is possible to solve explicitly for the gap given any value of $d$, the resulting expression is very complicated and we opt to simply plot the two-wire gap. All analytical expressions for the proximity-induced gaps in the asymmetric limit are plotted in Fig.~\ref{asymm}(b) choosing $k_Fx_w=\pi/4$.

\subsection{Strong Tunneling Limit}

The final limit that we consider numerically is that when tunneling between the superconductor and wires is strong, corresponding to $t_i^2/\mu_s\gg\Delta$. The numerical results are plotted in Fig.~\ref{strong}. In this limit, the proximity-induced gap is of the same order of magnitude as the bulk gap of the superconductor $E_g\sim\Delta$, except when the superconductor is very narrow ($d\ll\xi_s$). Furthermore, for $d\gg\xi_s$, the gap saturates to a value almost equal to the bulk gap $E_g\approx\Delta$ (we note in the numerical plot we never reach the limit $d\gg\xi_s$ due to our choice of small $\Delta$). The two-wire gap $E_g$, as well as the direct and crossed Andreev gaps, is plotted as a function of superconductor width in Fig.~\ref{strong}. The most notable difference when comparing to the weak-tunneling regime is that the drastic enhancement of the oscillation amplitude of the direct and crossed Andreev gaps in the limit $d\ll\xi_s$ is suppressed.

\begin{figure}[t!]
\includegraphics[width=0.45\linewidth]{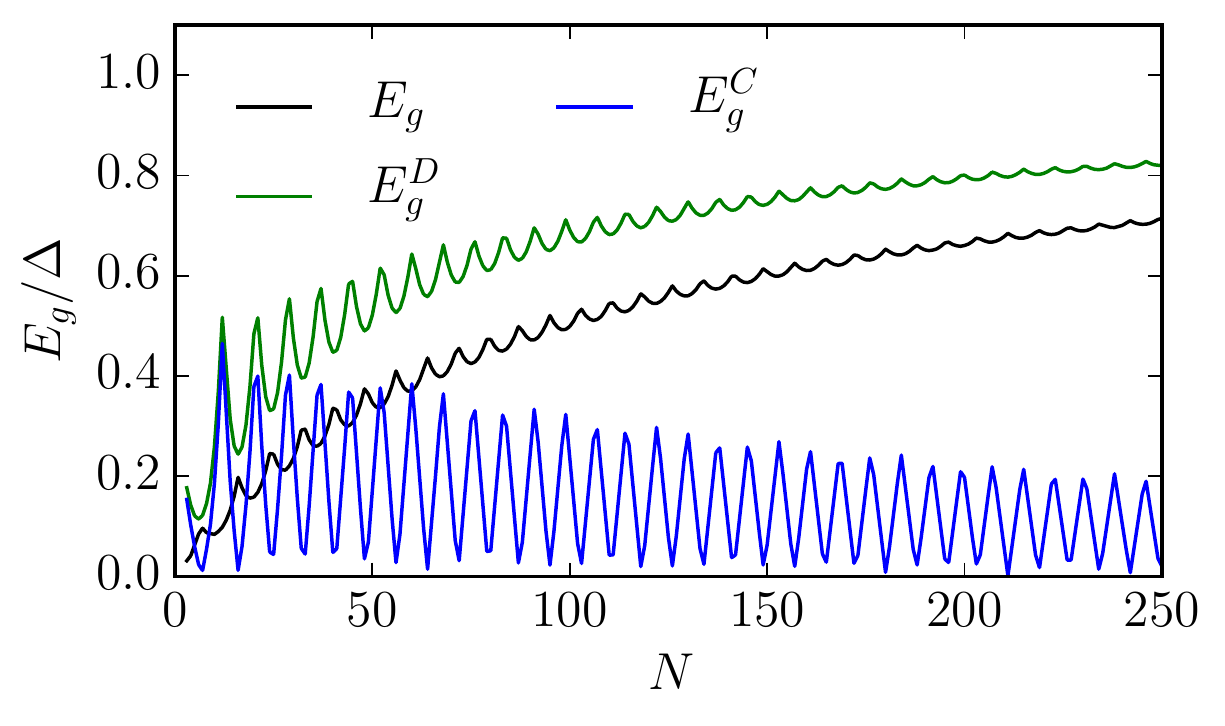}
\caption{\label{strong} Proximity-induced gaps calculated numerically within a tight-binding model when tunneling between superconductor and wires is very strong, plotted as a function of superconductor width $N$. We choose a parameter set $\Delta=0.003t$, $\mu_s=0.3t$, $\mu_L=\mu_R=0.1t$, and $t_L=t_R=0.2t$.}
\end{figure}

\section{Integrating Out Superconductor} \label{Sec5supp}
In this section, we follow the usual method of ``integrating out" the superconducting degrees of freedom from Hamiltonian \eqref{Hamiltoniansupp} in order to determine an effective pairing theory describing the superconductivity induced in the two-nanowire system. We will first carry out the calculation in the Green's function representation, and we will then carry out the calculation in the equivalent path integral representation. We will also show how this method can be used to reproduce our exact diagonalization and discuss the limitations of employing this method.
\subsection{Green's function representation} \label{GreensFunctionSec}
We begin with the same Hamiltonian as given in Eq.~\eqref{Hamiltoniansupp}. In the Heisenberg representation, the field operators in the system must obey the following imaginary-time equations of motion:
\begin{subequations} \label{eqsofmotionsupp}
\begin{gather}
\frac{\partial}{\partial\tau}\psi_{i\sigma}(y_i,\tau)=-\sum_{\sigma'}\left[\mathcal{H}_{\sigma\sigma'}(y_i)\psi_{i\sigma'}(y_i,\tau)+\int d\x\,T^{i\dagger}_{\sigma\sigma'}(y_i,\x)\psi_{s\sigma'}(\x,\tau)\right], \\
\frac{\partial}{\partial\tau}\psi_{i\sigma}^\dagger(y_i,\tau)=\sum_{\sigma'}\left[\mathcal{H}^{iT}_{\sigma\sigma'}(y_i)\psi_{i\sigma'}^\dagger(y_i,\tau)+\int d\x\,T^{iT}_{\sigma\sigma'}(y_i,\x)\psi_{s\sigma'}^\dagger(\x,\tau)\right], \\
\frac{\partial}{\partial\tau}\psi_{s\sigma}(\x,\tau)=-\sum_{\sigma'}\left[H_0(\x)\psi_{s\sigma}(\x,\tau)+\Delta_{\sigma\sigma'}\psi_{s\sigma'}^\dagger(\x,\tau)+\int dy_i\,T^{i}_{\sigma\sigma'}(\x,y_i)\psi_{i\sigma'}(y_i,\tau)\right], \\
\frac{\partial}{\partial\tau}\psi^\dagger_{s\sigma}(\x,\tau)=\sum_{\sigma'}\left[H_0(\x)\psi^\dagger_{s\sigma}(\x,\tau)-\Delta^\dagger_{\sigma\sigma'}\psi_{s\sigma'}(\x,\tau)+\int dy_i\,T^{i*}_{\sigma\sigma'}(\x,y_i)\psi^\dagger_{i\sigma'}(y_i,\tau)\right],
\end{gather}
\end{subequations}
We can define an $8\times8$ matrix (in wire $\otimes$ Nambu $\otimes$ spin space) Green's function describing states in the two wires as
\begin{equation}
{G}_{ij}(y_i,\tau;y_j',\tau')=-{\tau}_z\bigl\langle T_\tau\Psi_i(y_i,\tau)\Psi_j^\dagger(y_j',\tau')\bigr\rangle,
\end{equation}
where we introduce Nambu spinor $\Psi_i=[\psi_{i\uparrow},\psi_{i\downarrow},\psi^\dagger_{i\uparrow},\psi^\dagger_{i\downarrow}]^T$ (note that the Pauli matrix ${\tau}_z=\tau_z\otimes\sigma_0$ should not be confused with the imaginary time $\tau$). We can also define a superconducting Green's function by
\begin{equation}
{G}_s(\x,\tau;\x',\tau')=-\tau_z\bigl\langle T_\tau\Psi_s(\x,\tau)\Psi_s^\dagger(\x',\tau')\bigr\rangle.
\end{equation}
Finally, we define two mixed Green's functions by
\begin{subequations}
\begin{align}
{G}_{T_j}(\x,\tau;y_i',\tau')&=-\tau_z\bigl\langle T_\tau\Psi_s(\x,\tau)\Psi_j^\dagger(y_j',\tau')\bigr\rangle, \\
{G}_{T^\dagger_i}(y_i,\tau;\x',\tau')&=-\tau_z\bigl\langle T_\tau\Psi_i(y_i,\tau)\Psi_s^\dagger(\x',\tau')\bigr\rangle,
\end{align}
\end{subequations}
Using Eqs.~\eqref{eqsofmotionsupp}, we find that the equation of motion governing the wire Green's function is given by
\begin{equation} \label{wireGeq}
[i\omega-\check{\mathcal{H}}_i(y_i)]{G}_{ij}(y_i,y_j')-\int d\x\,\check{T}^\dagger_i(y_i,\x){G}_{T_j}(\x,y_j')=\delta_{ij}\delta(y_i-y_j'),
\end{equation}
where $\omega$ is a Matsubara frequency,
\begin{equation}
\check{\mathcal{H}}_i(y_i)=\begin{pmatrix} \mathcal{H}_i(y_i) & 0 \\ 0 & -\mathcal{H}_i^T(y_i) \end{pmatrix}
\end{equation}
is the Nambu-space Hamiltonian of wire $i$, and
\begin{equation}
\check{T}_i(\x,y_i)=\begin{pmatrix} T_i(\x,y_i) & 0 \\ 0 & -T^*(\x,y_i) \end{pmatrix}
\end{equation}
is a Nambu-space tunneling matrix. We now wish to express the mixed Green's function $G_{T_j}$ in Eq.~\eqref{wireGeq} in terms of the wire Green's function. The equation of motion for the mixed Green's function is given by
\begin{equation} \label{mixedGeq}
[i\omega-\check{\mathcal{H}}_\text{BCS}(\x)]{G}_{T_j}(\x,y_j')-\sum_i\int dy_{i1}\,\check{T}_i(\x,y_{i1}){G}_{ij}(y_{i1},y_j')=0,
\end{equation}
where
\begin{equation} \label{Ginvs}
\check{\mathcal{H}}_\text{BCS}(\x)=\begin{pmatrix}
	H_0(\x) & \Delta \\
	\Delta^\dagger & -H_0(\x)
	\end{pmatrix}
\end{equation}
is the Nambu-space Hamiltonian of the superconductor. We invert Eq.~\eqref{mixedGeq} to give
\begin{equation} \label{mixedGexp}
{G}_{T_j}(\x,y_j')=\sum_i\int dy_{i1}\int d\x_1\,{G}_s^0(\x,\x_1)\check{T}_i(\x_1,y_{i1}){G}_{ij}(y_{i1},y_j'),
\end{equation}
where the Green's function $G_s^0(\x,\x')$ satisfies
\begin{equation} \label{bareGeq}
[i\omega-\check{\mathcal{H}}_\text{BCS}(\x)]G_s^0(\x,\x')=\delta(\x-\x').
\end{equation}
Substituting Eq.~\eqref{mixedGexp} into Eq.~\eqref{wireGeq}, the equation for the wire Green's function is found to be
\begin{equation}
[i\omega-\check{\mathcal{H}}_{i}(y_i)]{G}_{ij}(y_i,y_j')-\int dy_{1}\,{\Sigma}_{ij}(y_i,y_{1}){G}_{ij}(y_{1},y_j')=\delta_{ij}\delta(y_i-y_j'),
\end{equation}
where we identify the self-energy induced on the two-nanowire system as
\begin{equation} \label{selfenergy}
{\Sigma}_{ij}(y_i,y_j')=\int d\x_1\int d\x_2\,\check{T}^\dagger_i(y_i,\x_1){G}_s^0(\x_1,\x_2)\check{T}_j(\x_2,y_j').
\end{equation}
This expression for the self-energy is analogous to those appearing in, for example, Refs.~[\onlinecite{Sau:2010prox,Kopnin:2011}], which deal with the self-energy induced in a single nanowire by a bulk superconductor. Diagonal elements of the self-energy ($i=j$) correspond to intrawire coupling terms while off-diagonal elements ($i\neq j$) correspond to interwire coupling terms.

\subsection{Path integral representation} \label{PathIntegralSec}
We again begin with the same Hamiltonian as in Eq.~(\ref{Hamiltoniansupp}). The action of the nanowires can be expressed as
\begin{equation}
S_{NW}^i=\frac{1}{2}\int\frac{d\omega}{2\pi}\int dy_i\,\Psi_i^\dagger(y_i)[i\omega-\check{\mathcal{H}}_i(y_i)](y_i)\Psi_i(y_i).
\end{equation}
Similarly, the action of the superconductor is given by
\begin{equation}
S_{BCS}=\frac{1}{2}\int\frac{d\omega}{2\pi}\int d\x\,\Psi_s^\dagger(\x)[i\omega-\check{\mathcal{H}}_\text{BCS}(\x)]\Psi_s(\x).
\end{equation}
Finally, the tunneling action is
\begin{equation}
S_t^i=\frac{1}{2}\int\frac{d\omega}{2\pi}\int dy_i\int d\x\left\{\Psi_i^\dagger(y_i)\check{T}_i^\dagger(y_i,\x)\Psi_s(\x)+\Psi_s^\dagger(\x)\check{T}_i(\x,y_i)\Psi_i(y_i)\right\}.
\end{equation}
The fermionic coherent state path integral representation for the partition function \cite{Altland} of this system is given by
\begin{equation} \label{partition}
\mathcal{Z}=\int D[\bar\psi_L,\psi_L]\int D[\bar\psi_R,\psi_R]\int D[\bar\psi_s,\psi_s]e^{-S_{NW}[\bar\psi_i,\psi_i]-S_{BCS}[\bar\psi_s,\psi_s]-S_t[\bar\psi_i,\psi_i,\bar\psi_s,\psi_s]},
\end{equation}
where $\bar\psi,\psi$ are the Grassman variables corresponding to the fermion fields $\psi^\dagger,\psi$. The path integral over superconducting fields is a Gaussian with exponent
\begin{equation}
\begin{aligned}
S_{BCS}+S_t&=\frac{1}{2}\int\frac{d\omega}{2\pi}\int d\x\biggl\{\bar\Psi_s(\x)[i\omega-\check{\mathcal{H}}_\text{BCS}(\x)]\Psi_s(\x) \\
	&+\bar\Psi_s(\x)\biggr[\sum_j\int dy_j\,\check{T}_j(\x,y_j)\Psi_j(y_j)\biggr]+\biggl[\sum_i\int dy_i\,\bar\Psi_i(y_i)\check{T}_i^\dagger(y_i,\x)\biggr]\Psi_s(\x)\biggr\}.
\end{aligned}
\end{equation}
The quadratic action can then be rewritten in the form
\begin{equation}
\begin{aligned}
S_{BCS}+S_t&=\frac{1}{2}\int\frac{d\omega}{2\pi}\int d\x\biggl[\bar\Psi_s(\x)+\sum_i\int dy_i\int d\x_1\,\bar\Psi_i(y_i)\check{T}^\dagger_i(y_i,\x_1){G}_s^0(\x_1,\x)\biggr][i\omega-\check{\mathcal{H}}_\text{BCS}(\x)] \\
	&\times\biggl[\Psi_s(\x)+\sum_j\int dy_j\int d\x_1\,{G}_s^0(\x,\x_1)\check{T}_j(\x_1,y_j)\Psi_j(y_j)\biggr] \\
	&-\frac{1}{2}\sum_{i,j}\int\frac{d\omega}{2\pi}\int dy_i\int dy_j\int d\x_1\int d\x_2\,\bar\Psi_i(y_i)\check{T}_i^\dagger(y_i,\x_1){G}_s^0(\x_1,\x_2)\check{T}_j(\x_2,y_j)\Psi_j(y_j)\biggr\},
\end{aligned}
\end{equation}
where $G_s^0(\x,\x')$ is as defined in Eq.~\eqref{bareGeq}. Evaluating the Gaussian path integral yields an effective action describing the two-nanowire system given by
\begin{equation} \label{newaction3}
S_\text{eff}=\frac{1}{2}\sum_{i,j}\int\frac{d\omega}{2\pi}\int dy_i\int dy_j\,\bar\Psi_i(y_i)\left\{\delta_{ij}\delta(y_i-y_j)[i\omega-\check{\mathcal{H}}_i(y_i)]-\int d\x_1\int d\x_2\,\check{T}^\dagger_i(y_i,\x_1){G}_s^0(\x_1,\x_2)\check{T}_j(\x_2,y_j)\right\}\Psi_j(y_j).
\end{equation}
Note that the second term within braces in Eq.~\eqref{newaction3} represents terms induced on the bare two-nanowire system by the superconductor; comparing with Eq.~\eqref{selfenergy}, this term is precisely the self-energy which was calculated in the Green's function representation.

\subsection{Integrating out with bulk Green's function in a finite geometry}
For the choice of tunneling matrix as given in Eq.~\eqref{tunnelingsupp}, the self-energy induced on the two-nanowire system by the superconductor is given by
\begin{equation} \label{selfenergy2}
\Sigma_{ij}(x_i,x_j)=t_it_j\tau_zG_s^0(x_i,x_j)\tau_z.
\end{equation}
Typically, the self-energy is evaluated with the bulk BCS Green's function, which is expressed in momentum space as
\begin{equation} \label{Gk}
G_s^0(k)=-\frac{i\omega+\xi_{ks}\tau_z+\Delta\tau_x}{\Delta^2+\xi_{ks}^2+\omega^2}.
\end{equation}
We define $\xi_{ks}=k^2/2m_s-\mu_s$, where $\kk=(k_x,k_y)$ is a two-dimensional momentum. 

In this section, we calculate the self-energy when using the bulk Green's function to integrate out the superconductor in a finite geometry. To justify the use of a translationally-invariant Green's function to describe a finite geometry, we necessarily impose periodic boundary conditions at $x=0$ and $x=d$, taking the wires to be located at the ends of the superconductor. This allows us to define the Fourier transform
\begin{equation}
G_s^0(x-x')=\frac{1}{d}\sum_{k_x}G_s^0(k)e^{ik_x(x-x')},
\end{equation}
where the momentum takes quantized values $k_x=2n\pi/d$ for $n\in\mathbb{Z}$. In real space, intrawire couplings correspond to $G_s^0(0)$ while interwire couplings correspond to $G_s^0(d)$. Performing the sum over $k_x$ in the limit $E\ll\Delta$, we find the intrawire terms
\begin{equation}
G_s^0(0)=-\frac{\tau_x\sinh\chi-\tau_z\sin(k_Fd)}{v_F[\cosh\chi-\cos(k_Fd)]}
\end{equation}
and the interwire terms
\begin{equation}
G_s^0(d)=-\frac{e^{-\chi/2}}{v_F[\cosh\chi-\cos(k_Fd)]}\bigl\{\tau_x[\sinh\chi-2\sin^2(k_Fd/2)]\cos(k_Fd/2)\bigr\}-\tau_z[\sinh\chi+2\cos^2(k_Fd/2)]\sin(k_Fd/2)\bigr\}.
\end{equation}
Given the Green's function in real space, we find an effective Hamiltonian describing the two-nanowire system given by
\begin{equation}
\mathcal{H}_\text{eff}=\begin{pmatrix}
	\xi_k+\beta\gamma_L & \alpha\gamma_L & \eta\sqrt{\gamma_L\gamma_R} & \delta\sqrt{\gamma_L\gamma_R} \\
	\alpha\gamma_L & -\xi_k-\beta\gamma_L & \delta\sqrt{\gamma_L\gamma_R} & -\eta\sqrt{\gamma_L\gamma_R} \\
	\eta\sqrt{\gamma_L\gamma_R} & \delta\sqrt{\gamma_L\gamma_R} & \xi_k+\beta\gamma_R & \alpha\gamma_R \\
	\delta\sqrt{\gamma_L\gamma_R} & -\eta\sqrt{\gamma_L\gamma_R} & \alpha\gamma_R & -\xi_k-\beta\gamma_R
	\end{pmatrix},
\end{equation}
where we define the quantities $\alpha=\sinh\chi/D$, $\beta=\sin(k_Fd)/D$, $\delta=e^{-\chi/2}[\sinh\chi-2\sin^2(k_Fd/2)]\cos(k_Fd/2)/D$, $\eta=e^{-\chi/2}[\sinh\chi+2\cos^2(k_Fd/2)]\sin(k_Fd/2)/D$, and $D=\cosh\chi-\cos(k_Fd)$.

Expanding in the limit $\chi\ll1$ ($d\ll\xi_s$) to lowest order, we can replace $\alpha=0$, $\beta=\cot(k_Fd/2)$, $\delta=-\cos(k_Fd/2)e^{-\chi/2}$, and $\eta=\cos(k_Fd/2)\cot(k_Fd/2)e^{-\chi/2}$. For simplicity, we take $\gamma_L=\gamma_R=\gamma_d$ and $\sqrt{\gamma_L\gamma_R}=\gamma_c$; although $\gamma_d=\gamma_c$, we differentiate terms originating from intrawire and interwire processes with the subscripts $d$ and $c$. In this limit, we find a spectrum given by
\begin{equation}
E_\pm^2(\xi_k)=\delta^2\gamma_c^2+(\beta\gamma_d+\xi_k\pm\eta\gamma_c)^2.
\end{equation}
The excitation gap of this system is given by
\begin{equation}
E_g=|\delta|\gamma_c=\gamma_c|\cos(k_Fd/2)|
\end{equation}

In the opposite limit of a very wide superconductor ($\chi\gg1$), we replace $\alpha=1$, $\beta=\delta=\eta=0$ and find two branches of the excitation spectrum given by $E^2=\gamma_{L(R)}^2+\xi_k^2$. In this limit, integrating out using the bulk Green's function in a finite geometry does give the correct result, a consequence of the fact that the boundary effects that were not treated properly become unimportant in this limit.

\subsection{Properly accounting for boundary conditions}
We showed in the previous section that the bulk Green's function cannot be used to evaluate the self-energy in a finite geometry. As discussed in the main text, this is a result of not properly accounting for the boundary conditions of the system. In this section, we instead opt to evaluate the self-energy using the Green's function of a finite-sized superconductor which obeys vanishing boundary conditions at both $x=0$ and $x=d$.

The Green's function of a bulk superconductor expressed in real space is given by
\begin{equation} \label{bulkG}
G_\text{bulk}(x-x')=-\frac{1}{v_F\Omega}\left[(i\omega+\Delta\tau_x)\cos(k_F|x-x'|)-\Omega\tau_z\sin(k_F|x-x'|)\right]e^{-\Omega|x-x'|/v_F},
\end{equation}
where $\Omega=\sqrt{\Delta^2+\omega^2}$ and, as we have done throughout, we approximate $p_\pm=k_F\pm i\Omega/v_F$. We note that this bulk Green's function is different than the one calculated in the previous section because here we Fourier transform Eq.~\eqref{Gk} over continuous $k_x$. The bare Green's function which obeys the appropriate vanishing boundary conditions is given by
\begin{equation} \label{bareGsol}
\begin{aligned}
G_s^0(x,x')&=\frac{1}{2v_F\Omega}(i\omega+\Delta\tau_x+i\Omega\tau_z)\left\{\frac{\sin[p_+(d-x')]}{\sin(p_+d)}e^{ip_+x}+[i+\cot(p_+d)]\sin(p_+x')e^{-ip_+x}\right\} \\
	&+\frac{1}{2v_F\Omega}(i\omega+\Delta\tau_x-i\Omega\tau_z)\left\{[-i+\cot(p_-d)]\sin(p_-x')e^{ip_-x}+\frac{\sin[p_-(d-x')]}{\sin(p_-d)}e^{-ip_-d}\right\}+G_\text{bulk}(x-x').
\end{aligned}
\end{equation}
Substituting Eq.~\eqref{bareGsol} into the self-energy Eq.~\eqref{selfenergy2} while assuming $\gamma,E\ll\Delta$ and $x_w\ll d$, we find intrawire terms
\begin{equation}
\Sigma_{ii}(x_i,x_i)=\frac{\gamma_i}{\cosh(2\chi)-\cos(2k_Fd)}\biggl[2\sinh(2\chi)\sin^2(k_Fx_w)\tau_x+\{\sin(2k_Fd)-\cosh(2\chi)\sin(2k_Fx_w)-\sin[2k_F(d-x_w)]\bigr\}\tau_z\biggr].
\end{equation}
The term proportional to $\tau_x$ corresponds precisely to the direct pairing $\Delta_i$ that we obtained by mapping our exact solution to an effective pairing model [Eq.~\eqref{deltaDeff}], while the term proportional to $\tau_z$ corresponds to the chemical potential shift $\delta\mu_i$ [Eq.~\eqref{deltamueff}]. We neglect the term of the self-energy proportional to $\tau_0$, as it is proportional to $E/\Delta\ll1$. Additionally, we find the interwire terms
\begin{equation}
\Sigma_{i\bar i}(x_i,x_{\bar i})=\frac{4\sqrt{\gamma_L\gamma_R}\sin^2(k_Fx_w)}{\cosh(2\chi)-\cos(2k_Fd)}\biggl[\sinh\chi\cos(k_Fd)\tau_x+\cosh\chi\sin(k_Fd)\tau_z\biggr].
\end{equation}
The term proportional to $\tau_x$ corresponds precisely to the crossed Andreev pairing strength $\Delta_c$ [Eq.~\eqref{deltaCeff}], while the term proportional to $\tau_z$ corresponds to the interwire tunneling strength $\Gamma$ [Eq.~\eqref{Gammaeff}]. We thus find that the results of our exact diagonalization can be reproduced by appropriately choosing the bare superconducting Green's function with which to evaluate the self-energy of Eq.~\eqref{selfenergy2}.

\subsection{Integrating out with bulk Green's function in an infinite geometry} \label{EffectiveIntOutSec}
While the bulk Green's function cannot be used to integrate out the superconductor in a finite geometry, we note that it can be used assuming that the superconductor is infinitely large. Such a situation is realized in the geometry shown in Fig.~\ref{setup2}, where two nanowires (with interwire separation $d$) are placed on top of an infinite 2D superconducting plane.

\begin{figure}[t!]
\includegraphics[width=0.3\linewidth]{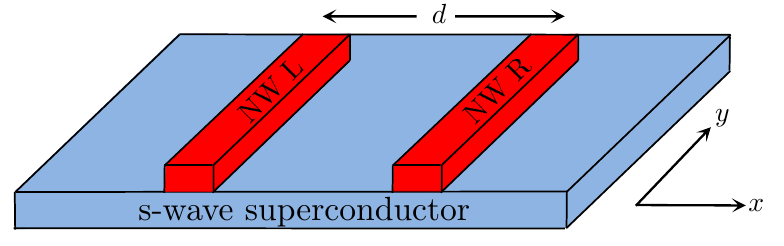}
\caption{\label{setup2} Alternate geometry considered. Two infinite nanowires are placed on top of an infinite 2D superconducting plane and are separated by a distance $d$.}
\end{figure}

Substituting the bulk Green's function Eq.~\eqref{bulkG} into the self-energy Eq.~\eqref{selfenergy2}, we find the intrawire term
\begin{equation}
\Sigma_{ii}(x_i,x_i)=\gamma_i\tau_x
\end{equation}
and interwire terms
\begin{equation}
\Sigma_{i\bar i}(x_i,x_{\bar i})=\sqrt{\gamma_L\gamma_R}\left[\tau_x\cos(k_Fd)e^{-\chi}+\tau_z\sin(k_Fd)e^{-\chi}\right].
\end{equation}
We can therefore identify an effective pairing Hamiltonian describing the two-wire system which is the same as in Eq.~\eqref{effectivemodelH}, with the effective parameters
\begin{equation}
\begin{gathered}
\Delta_i=\gamma_i, \\
\delta\mu=0, \\
\Delta_c=\sqrt{\gamma_L\gamma_R}\cos(k_Fd)e^{-\chi}, \\
\Gamma=-\sqrt{\gamma_L\gamma_R}\sin(k_Fd)e^{-\chi}.
\end{gathered}
\end{equation}
In the new geometry, single-particle intrawire tunneling is completely suppressed while the direct pairing gap becomes independent of $d$ (this should be obvious, because if we decouple one of the wires then the quantity $d$ no longer has any physical meaning). Notably, we still always have $\Delta_c^2<\Delta_L\Delta_R$.

We now utilize this mapping to determine the excitation gap in the system. For simplicity, let us assume that tunneling is symmetric ($\gamma_L=\gamma_R$) so that $\Delta_L=\Delta_R=\Delta_d$. In this case, we obtain an excitation spectrum given by
\begin{equation}
E^2_\pm=(\xi_k\mp\Gamma)^2+(\Delta_d\pm\Delta_c)^2,
\end{equation}
which yields an excitation gap
\begin{equation} \label{infinitegapsupp}
E_g(d)=\Delta_d-|\Delta_c|=\gamma\left(1-e^{-\chi}|\cos (k_Fd)|\right).
\end{equation}
Similarly, we can find the crossed Andreev gap corresponding to the quantum Hall regime by setting $\Gamma=\Delta_i=0$ in the effective Hamiltonian [Eq.~\eqref{effectivemodelH}]. We then find an excitation spectrum given by $E^2=\xi_k^2+\Delta_c^2$ and a gap
\begin{equation} \label{infiniteCAgapsupp}
E_g^C(d)=|\Delta_c|=\gamma e^{-\chi}|\cos (k_Fd)|.
\end{equation}

We plot the full and crossed Andreev gaps obtained analytically in Fig.~\ref{torus}(a). To test our analytical calculations, we again run a numerical tight-binding calculation on the geometry of Fig.~\ref{setup2}. For computational reasons and to avoid boundary effects, we implement the infinite geometry by considering the bulk two-dimensional superconductor to be a torus (corresponding to imposing periodic boundary conditions). We take the circumferential length $La_x$ ($a_x$ is the lattice constant and $L$ is the number of sites) of the torus to be much longer than the superconducting coherence length. The wires are tunnel-coupled to two superconducting sites separated by the distance $d=Na_x$. The results of the numerical calculation are plotted in Fig.~\ref{torus}(b), showing good agreement with the analytical calculation.

\begin{figure}[t!]
\centering
\includegraphics[width=\linewidth]{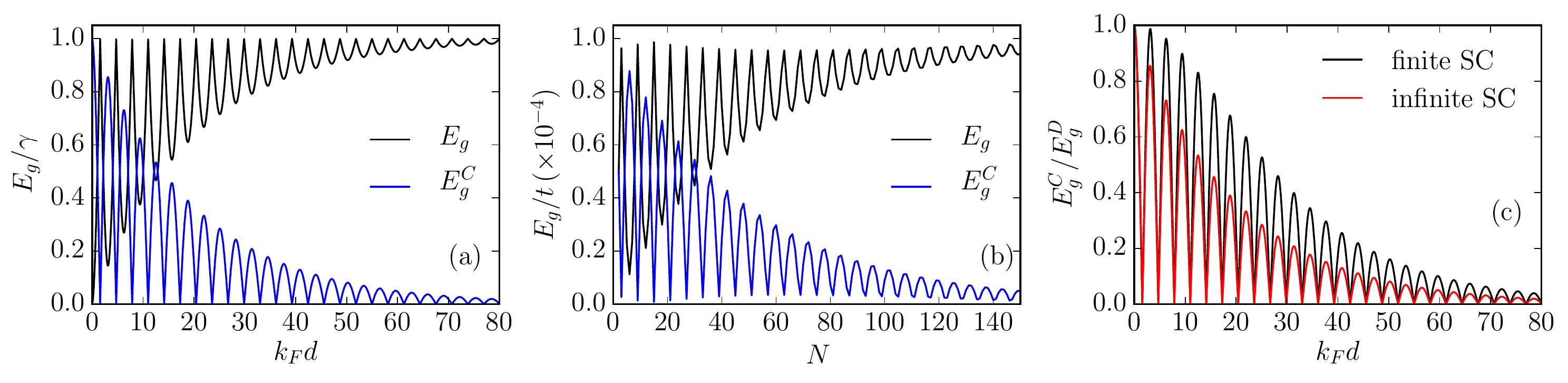}
\caption{\label{torus} (a) Proximity-induced gaps obtained analytically for the geometry of two nanowires coupled to an infinite superconducting strip. The black curve corresponds to symmetric tunneling ($\gamma_L=\gamma_R$) [Eq.~\eqref{infinitegapsupp}] and the blue curve corresponds to the quantum Hall regime ($\gamma_{L\uparrow}=\gamma_{R\downarrow}\neq0$ and $\gamma_{L\downarrow}=\gamma_{R\uparrow}=0$) [Eq.~\eqref{infiniteCAgapsupp}]. (b) Proximity-induced superconducting gaps obtained numerically in torus geometry (with circumferential length $L=600$), plotted as a function of interwire separation $Na_x$ for parameters $\Delta=0.02t$, $\mu_s=0.3t$, $\mu_n=0.03t$. This corresponds to modeling two wires placed on the top of an infinite superconducting strip. Black curve corresponds to two-wire case ($t_L=t_R=0.01t$) and blue curve corresponds to quantum Hall case ($t_{L\uparrow}=t_{R\downarrow}=0.01t$ and $t_{L\downarrow}=t_{R\uparrow}=0$). (c) Ratios of the crossed Andreev gap $E_g^C$ to the direct gap $E_g^D$ in the two geometries considered in this paper [Eqs.~\eqref{ratiofinite}-\eqref{ratioinfinite}].}
\end{figure}

It is also useful to compare the ratio $\Delta_c/\Delta_d$ in the two geometries considered in this paper. First, in the finite geometry considered in previous sections, this ratio is given by (see Sec.~\ref{Sec3supp} for explicit forms of the effective pairing parameters)
\begin{equation} \label{ratiofinite}
\left(\frac{\Delta_c}{\Delta_d}\right)_\text{finite}=\frac{2\sinh\chi\cos (k_Fd)}{\sinh(2\chi)}.
\end{equation}
In the infinite geometry, on the other hand, we have
\begin{equation} \label{ratioinfinite}
\left(\frac{\Delta_c}{\Delta_d}\right)_\text{infinite}=e^{-\chi}\cos (k_Fd).
\end{equation}
This ratio is plotted for the two different cases in Fig.~\ref{torus}(c). We see that the ratio is always larger in the finite geometry, which can also be seen explicitly by considering the ratio
\begin{equation}
\left(\frac{\Delta_c}{\Delta_d}\right)_\text{finite}\biggl/\left(\frac{\Delta_c}{\Delta_d}\right)_\text{infinite}=\frac{2}{e^{-2\chi}+1}\geq1.
\end{equation}
Because one typically would want the ratio of the crossed Andreev pairing to the direct pairing to be as large as possible, the finite geometry seems to work a bit better for achieving this.

\newpage

\end{widetext}

\end{document}